\documentstyle[12pt]{article}

\input{epsf}

\def\dss#1{\displaystyle{#1}}
\def\omegr{\omega_{\bf R}}
\def\omegc{\omega_{\bf C}}
\def\Omegr{\Omega_{\bf R}}
\def\Omegc{\Omega_{\bf C}}
\def\lc{\Lambda_{\bf C}}
\def\wc{W_{\bf C}}

\def\tsix{{\bf T}^6}
\def\tfour{{\bf T}^4}
\def\dualtfour{\widetilde{\bf T}^4}
\def\ttwo{{\bf T}^2}

\def\cwp{{\cal W}_+}
\def\cwm{{\cal W}_-}

\def\trgt{\widehat{\cal M}}

\def\homegr{\hat{\omega}_{\bf R}}

\def\trace{\mbox{tr}\hspace{1pt}}

\begin{document}

\begin{titlepage}
\begin{flushright}
PUPT-1878\\
ITEP-TH-32/99\\
CALT-68-2241
\end{flushright}

\begin{center}
{\Large $ $ \\ $ $ \\
D1D5 System and Noncommutative Geometry}\\
\bigskip\bigskip\bigskip
{\large Andrei Mikhailov}
\footnote{On leave from the Institute of Theoretical and 
Experimental Physics, 117259, Bol. Cheremushkinskaya, 25, 
Moscow, Russia.}\\
\bigskip
California Institute of Technology, Pasadena, CA 91125\\
and\\
Department of Physics, Princeton University\\
Princeton, NJ 08544, USA\\
\vskip 1cm
E-mail: andrei@viper.princeton.edu
\end{center}
\vskip 1cm
\begin{abstract}
Supergravity on $AdS_3\times S^3\times {\bf T}^4$ has a dual description
as a conformal sigma-model with the target space being the moduli space of
instantons on the noncommutative torus. We derive the precise relation
between the parameters of this noncommutative torus and the parameters
of the near-horizon geometry. We show that the low energy dynamics
of the system of $D1D5$ branes wrapped on the torus of finite
size is described in terms of the noncommutative geometry. As a byproduct, 
we give a prediction
on the dependence of the moduli space of instantons on the noncommutative
${\bf T}^4$ on the metric and the noncommutativity parameter.
We give a compelling evidence that the moduli space of stringy instantons 
on ${\bf R}^4$ with the $B$ field does not receive $\alpha'$-corrections.
We also study the relation between the $D1D5$ sigma-model instantons and the 
supergravity instantons.  
\end{abstract}
\end{titlepage}

\section{Introduction.}

The formalism of noncommutative geometry is useful in string theory
when one studies certain special points in the moduli space of
compactifications. In string theory, small size manifolds are 
related to finite size manifolds by T duality.
However, the required T duality transformation usually does
not have a nice limit when the size of the manifold goes to zero.
The form of the T duality which brings the background with the small size
manifold to the background with the finite size manifold depends very 
irregularly on the moduli of the small size manifold, such as the metric 
and the $B$ field.
Therefore, one could naively expect that the physics also does not have
a regular limit. But it turns out that there exists a smooth
description of physics at very small distances in terms of the noncommutative
geometry. This amazing fact was discovered in string perturbation theory
in \cite{DH} and in the study of the Matrix Theory in \cite{CDS}.
The relation between the noncommutative geometry and the string theory
has been rigorously derived and made more precise in 
\cite{EdnaMorten,ChuHo,Schomerus,AASJ}. 
The authors of \cite{SWNC} have found the zero slope limit in
which  the noncomutative Yang-Mills
theory becomes a valid description of the open string sector of
the string theory, 
established the relation between the conventional and the noncommutative
gauge fields and found the stringy interpretation of the Morita
equivalence \cite{Rieffel}. Unfortunately, all the successfull
applications of noncommutative geometry to string theory have so far
been limited to flat backgrounds, although the natural
prescription for constructing the noncommutative algebra from the
background with the $B$-field is known in mathematical literature
\cite{Kontsevich,CF}.

Putting a field theory on the noncommutative space effectively introduces
the non-locality. This non-locality should
presumably lead to softening some of the field theory divergences
(as discussed, for example, in \cite{TwoLectures}).
One very important example of divergences affected by the noncommutativity
is the divergences in perturbation theory; the study of the effects
of noncommutativity in Feynman diagramms has been initiated in
\cite{Filk}. The other example, directly related to our paper, 
is the resolution of singularities in the instanton
moduli spaces. It was pointed out in \cite{NS} that introducing 
noncommutativity for the field theory on ${\bf R}^4$ is equivalent
from the point of view of the instanton moduli space to considering
the ADHM construction with the nonzero value of the 
Hyper-K\"ahler moment map.
(In other words, it introduces the Fayet-Illiopoulos terms in the
ADHM sigma-model). This discovery has an important application in the 
study of the string theory on Anti-deSitter space. It was conjectured
in \cite{Maldacena}, that this theory has a dual description in terms
of the conformal field theory. More precisely, the Type IIB string theory
on $AdS_3\times S^3\times X$, where $X$ is $K3$ or ${\bf T}^4$, is
dual to the conformal sigma-model whose target space is the moduli space
of instantons on $X$. This proposal, suggested in \cite{Maldacena}
and further studied in \cite{GKS,deBoerVir,KS},
is valid at the specific subspace of the moduli space. On this subspace,
the fluxes of the NSNS $B$ field through the two-cycles of $X$ are zero.
It was found in \cite{ABS}, that the non-zero fluxes of $B$
correspond to the non-zero value of the moment map in the ADHM construction.
Therefore, the string theory on $AdS_3\times S^3\times X$ with the nonzero $B$ field 
corresponds to the conformal sigma-model on the moduli space of instantons on
the {\em noncommutative} $X$. 

The purpose of this paper is to fix the precise relation between
the parameters of the noncommutative SYM theory and the parameters
of the near-horizon geometry. 
We will proceed in the following way. Wrapping $Q_4$ $D5$ branes on
$X$, $Q_2^i$ $D3$ branes on the $i$th two-cycle of $X$, and
adding $Q_0$ D-strings unwrapped, we end up having a black string
in the $5+1$-dimensional flat space. 
The background $AdS_3\times S^3\times X$ 
arises in Type IIB as the near-horizon limit of this black string.
The corresponding supergravity solution looks in a neighborhood 
of a given point 
in the $5+1$-dimensional space  as ${\bf R}^{5+1}\times X$, 
but the moduli of $X$ 
depend on the distance from the black string. For such a solution, let us consider
the geometry of $X$ near the horizon (which we denote $X^h$) and $X$ far away 
from the string (we denote it $X^{\infty}$). 
Given $X^{\infty}$, one can find $X^h$, using the supergravity equations of motion.
It turns out that $X^h$ cannot be arbitrary $K3$ (or ${\bf T}^4$);
the moduli of $X^h$ should satisfy certain attractor conditions
\cite{Attractors}. On the other hand, the moduli of $X^{\infty}$ can
be arbitrary. Therefore, the correspondence $X^{\infty}\mapsto X^h$ is 
not one-to-one; there is a whole family of $X^{\infty}$ flowing to a given $X^h$.

The CFT dual to the near-horizon supergravity can be found using the
procedure suggested by Maldacena \cite{Maldacena}. We look at the worldsheet theory
of the corresponding system of branes. In the low energy limit,
it becomes a conformal theory. According to the Maldacena conjecture, this is the
conformal theory dual to $AdS$ supergravity. The worldsheet effective
theory feels the flat background, the one which would be there without branes.
The geometry of this background is ${\bf R}^{1+5}\times X^{\infty}$. Therefore
the moduli of the conformal theory depend on $X^{\infty}$. 
However, as we have already mentioned there are many possible $X^{\infty}$ which
give the same near-horizon $X^h$. The low energy theories on the worldvolume
of the black string should be equivalent for all the backgrounds
$X^{\infty}$ flowing to the same $X^h$. This is a necessary condition for $AdS/CFT$
correspondence to be self-consistent. It was argued in \cite{Maldacena}, that the
equivalence of the conformal theories corresponding to different $X^{\infty}$ 
flowing to the same $X^h$ follows from supersymmetry. 

We want to ask the following question: given the family of flat backgrounds 
$X^{\infty}$ flowing to the same near-horizon background, can we find a
"canonical" representative,  the one 
which gives the simplest description of the $D1D5$ worldsheet theory?
At first sight, the most appealing possibility would be to find 
$X^{\infty}$ of a very large size. 
If we could do that, we would get the description of
our CFT as the sigma model with the target space being 
the moduli space of Yang-Mills instantons. 
But it turns out that most of the attractors $X^h$ cannot be obtained as 
the near-horizon 
limit of the solutions with very large $X^{\infty}$. (In fact, there is 
an upper limit 
on the volumes of $X^{\infty}$ flowing to a given $X^h$). What we can do instead
is to find a representative with very {\em small} $X^{\infty}$.
For such backgrounds the low-energy effective action is formulated
rather explicitly in terms of the noncommutative geometry. We can use the results 
of \cite{SWNC}
relating the shape of $X^{\infty}$ and the $B$ field fluxes to the parameters
of the noncommutative manifold which we will denote $\widetilde{X^{\infty}}$.
According to \cite{SWNC}, 
the effective low-energy theory for the bound state 
of $D1$, $D3$ and $D5$ is supersymmetric Yang-Mills theory on 
$\widetilde{X^{\infty}}\times{\bf R}^{1+1}$. Therefore,
the effective low-energy theory for the black string is $1+1$-dimensional
sigma-model with the target space the moduli space of instantons 
on $\widetilde{X^{\infty}}$. (In fact, there is a  subtlety here which we will 
discuss later). This gives the dual description of the
near-horizon supergravity.

Let us mention an interesting consequence of this correspondence.
It turns out that even if we restrict ourselves to
the small-size $X^{\infty}$,
there is still the whole family of them flowing to the same $X^h$.
This gives us the family of noncommutative manifolds having the same moduli space
of instantons. Indeed, the corresponding sigma-models should be equivalent;
the equivalence of $(4,4)$ sigma-models implies that they have the same target 
spaces, modulo some discrete symmetries; but the equivalence classes
turn out to be connected, therefore the possibility of discrete symmetries 
is excluded. 
It has been proven in \cite{SWNC} that instantons on ${\bf R}^4$ depend
only on the self-dual part of the noncommutativity parameter. Our method allows us 
to formulate the analogue of this statement for instantons on ${\bf T}^4$.

The relation between the near-horizon geometry and the moduli of
the dual CFT has been studied in great detail in the paper by R.~Dijkgraaf 
\cite{Dijkgraaf}. The precise correspondence between the parameters of the attractor 
and the moduli space of the dual CFT has been given in that paper. Combined 
with our paper, this gives the description of the geometry of the moduli space 
of instantons on the noncommutative torus. 

Another application of our formulas is the study of the low-energy dynamics
of D-branes wrapped on the torus of the finite size (of the order $l_s$).
The most straightforward way of doing it would consist of two steps: 
1) study the low-energy worldvolume theory of $D5$ branes wrapped on
${\bf T}^4$ (with the numbers of $D1$ and $D3$ specifying the topological
sector of this theory); 2) compactify this low energy theory on ${\bf T}^4$,
to get the low energy theory of the black string. However, we think
that performing the first step is very hard for the finite size torus.
We do not know how to construct the low-energy theory describing the
slow degrees of freedom of $N>1$ branes of finite size in the nontrivial
topological sector. (The Nonabelian Born-Infeld Theory \cite{Tseytlin} 
is valid when the covariant derivatives of the field strength are small,
and this condition is not satisfied in our case).
But we can bypass the first step and go directly to the low energy dynamics 
of the black string. Indeed, our considerations show that
it is equivalent to the low energy dynamics of branes wrapped on
some very small torus. 
Therefore, the low energy Lagrangian for the black string can always be
represented as the sigma-model on the moduli space of noncommutative
instantons.

There is a subtlety in the relation between the instanton sigma model
and the low-energy supergravity on $AdS_3$. The statement that the low-energy
dynamics of the $D1D5$ system is described by the sigma-model on
the moduli space of instantons follows from the dimensional reduction
of the $1+5$-dimensional classical Super Yang-Mills theory on the 
noncommutative torus. Strictly speaking, we can treat this $1+5$-dimensional
theory classically only if it is weakly coupled on the compactification
scale. On the other hand, classical $AdS$ supergravity is valid only if
the radius of curvature of $AdS$ space is large enough. It turns out that
these two conditions are incompatible: we cannot trust dimensional
reduction in the regime where $AdS$ supergravity is valid. However,
we conjecture that the {\em shape} of the target-space of the sigma-model
describing the dynamics of the $D1D5$ system is independent of
the string coupling constant. We give two arguments confirming this
conjecture. The first argument is based on considering the dependence
of the sigma-model for the $D1D5$ system on the parameters of the background.
Our considerations in Section 3 will imply that the structure of equivalence
classes of backgrounds
giving the same sigma-model does not depend on $g_{str}$. Although 
this fact does not necessarily
 imply that the structure of the target space 
does not dependent of $g_{str}$,
we think that it supports the conjecture.  The second argument is based on
the relation between the world-sheet instantons of the $D1D5$ sigma
model and the D-instantons of the six-dimensional supergravity. 
An example of this relation was given in \cite{KoganLuzon}.
Instantons in toroidally compactified
string theories where considered in \cite{Pioline,PiolineKiritsis,Ganor}.
We argue that under certain conditions there is a correspondence between
these two kinds of instantons, and the action of the world-sheet instantons
is equal to the action of the supergravity instantons. The action of
the worldsheet instantons is related to the Hyper-K\"ahler 
period map of the target 
space. And the action of the supergravity 
instantons can be found when $g_{str}$ is small from supergravity.
The formula agrees with the period map for the target space conjectured
in \cite{Dijkgraaf} and does not contain $N$. This suggests
that the target space for small $g_{str}N$ and large $g_{str}N$ should
have the same period map. Therefore the Hyper-K\"ahler structure
of the target space is the same in the regime when we can use dimensional 
reduction and in the regime when we can trust supergravity. 

It was claimed in \cite{CGKM} that the moduli space of the twisted little
string theories on ${\bf T}^3$ is the moduli space of noncommutative 
instantons on ${\bf T}^4$. This result was derived by relating the
theory on $NS5$ brane to the $D2D6$ system, with the $D6$-brane wrapped on the
four-torus. Strictly speaking, $D2D6$ system on ${\bf T}^4$ is described by the
noncommutative Yang-Mills theory only when this four-torus
is very small. However, the situation with the $D2D6$ system should be
similar to what happens with $D1D5$: namely, the moduli space does
not depend on some combination of the background fields, and
there is a small torus in each equivalence class. Therefore, the moduli
space is always the moduli space of noncommutative instantons.

The paper is organized as follows.
In the second section, we give a brief review of the attractor equations
and explain when the supergravity approximation can be trusted.
In the third section, we give the argument for independence of the
sigma-model on certain combinations of the background fields, based on
supersymmetry. This section is auxiliary, and it is not necessary for
understanding the rest of the paper. Our arguments based on supersymmetry should be
closely related to the argument given in \cite{MaldacenaLE}, but we have tried
to make them more precise.
In the fourth section, we derive the relation between the asymptotic
background and the near-horizon (attractive) background.
This section is somewhat technical.  
The main formula is (\ref{WhichAreEquivalent}), giving the condition
for the two backgrounds to flow to the same attractor. The formula
(\ref{CiH}) expresses the near-horizon Ramond-Ramond fields in terms
of the Ramond-Ramond fields at infinity.
In the fifth section, we review the correspondence between the
string theory on the small torus and the Yang-Mills theory on the
noncommutative torus. Combining these results with the results from the
fourth section, we construct the flows on the moduli space of the 
noncommutative tori, which leave the moduli space of instantons invariant.
In Subsection 5.2 we make an observation which suggests that the moduli
space of instantons on noncommutative ${\bf R}^4$ does not receive
$\alpha'$-corrections.  
In the sixths section we discuss the relation between the world-sheet
instantons and the supergravity instantons.
In Appendix A, we review the correspondence between the moduli of the 
Type IIB compactification and the points of Grassmanian. 

\section{Attractor Equation.}
\subsection{Supergravity Solution.}
The dependence of the moduli of K3 or $T^4$ 
on the distance from the
horizon is given by the ``attractor equation''. This equation 
can be found in many papers, for example in 
\cite{dAuriaFre}, although in somewhat implicit form. 
In this section, we will briefly review the attractor equations and
present them in the form most convenient for our purposes.

We will concentrate on the case $X=K3$.
First of all, we want to review  some properties of the chiral $N=4$ 
$D=6$ supergravity. We will follow the original paper \cite{Romans}
but use slightly different language.
This theory describes an interaction of $N=4b$ gravity
multiplet with $n$ tensor multiplets. For the compactification
of Type IIB on $K3$, we have $n=21$. 
The gravity multiplet contains a graviton, four left-handed 
gravitini and five antisymmetric tensors with selfdual 
field strength. The tensor multiplet consists of an antisymmetric
tensor with antiselfdual field strength, four right-handed
symplectic Majorana spinors and five scalars. 
To describe the interacting theory, we think of the scalars as 
parametrizing the coset $$Gr(5,5+21)={SO(5,21)\over SO(5)\times SO(21)}$$
We represent this coset
as the manifold of five-dimensional positive planes in the space
$$L={\bf R}^{5+21}$$
For a given five-plane $W\subset L$, we denote $P_+$ the projector
on this plane, and $P_-$ the projector on the orthogonal plane.
We denote the $1+5$-dimensional Minkowski space $M$.
The field strengths of the gravity multiplet and the tensor multiplets
may be combined in a vector $H\in \Omega^3 M\otimes L$,
with the constraint
\begin{equation}
*H=(P_+-P_-)H
\end{equation}
Instead of writing $P_+H$ and $P_-H$, we will often write $H_+$ and $H_-$.
Let $S_-^{1/2}M$ be the bundle of antichiral
spinors on $M$, and $S_+^{3/2}M$ the bundle of Rarita-Schwinger fields
({\em i.e.,} chiral spinors with vector indices).
Also, we denote ${\cal W}_+$ the ``tautological''
bundle on the Grassmanian $Gr(5,5+21)$, that is
the bundle whose fiber over the point represented by the plane $W$ is the
plane $W$ itself. Similarly,  ${\cal W}_-$ will denote the bundle whose 
fiber is $W^{\perp}$. The gravitino $\psi_{\mu}$ and the fermions $\chi$ 
from the tensor multiplets are the sections of the following bundles:
\begin{equation}
\begin{array}{l}\dss{
\psi\in\Gamma([S_+^{3/2}M\otimes S(\cwp)]_{\bf R})}\\
\dss{\chi\in\Gamma([S_-^{1/2}M\otimes S(\cwp)]_{\bf R}\otimes\cwm)}
\end{array}
\end{equation}
Here $S(\cwp)$ is the spinor bundle associated with the $Spin(5)$ vector
bundle\footnote{The moduli space of scalars is topologically trivial
(the global coordinates may be found, for example, in \cite{AAFL}).
Therefore, there is no difference 
between the $Spin(5)$ bundles and the $SO(5)$ bundles.}
$\cwp$ (we will call its sections ``internal spinors''). 
The subindices $\bf R$ mean that some Majorana 
conditions are imposed. Let us explain the Majorana conditions for $\psi$.
$(S_+^{3/2}M\otimes S(\cwp))_{\bf R}$ is generated
by the expressions of the form $\sum_a s_a\otimes\phi_a$, where 
$s_a\in S_+^{3/2}M$, $\phi_a\in S(\cwp)$, with the reality condition 
$$
\sum_a s_a\otimes\phi_a=\sum_a 
C_{(1)}\bar{s}^T_a\otimes C_{(2)}\bar{\phi}^T_a
$$
We have denoted $C_{(1)}$ and $C_{(2)}$ the charge conjugation matrices 
for the space-time spinors and the internal spinors, respectively.
The Majorana condition for $\chi$ is defined in the same way. 
The supersymmetry generators $\epsilon$ are space-time and internal spinors:
\begin{equation}
\epsilon\in\Gamma([S_+^{1/2}M\otimes S(\cwp)]_{\bf R})
\end{equation}
The supersymmetry transformations for the gravitino and $\chi$ are:
\begin{equation}
\begin{array}{c}
\dss{
\delta_{\epsilon}\psi_{\mu}=D_{\mu}\epsilon+
{1\over 4}\kappa_0 (H_{+\mu\nu\rho})
.(\gamma^{\nu\rho}\epsilon),
}\\
\dss{
\delta_{\epsilon}\chi=-{1\over 2}\gamma^{\mu}(\partial_{\mu}\phi).\epsilon
-{1\over 12}\kappa_0 \gamma^{\mu\nu\rho}H_{-\mu\nu\rho}\otimes\epsilon
}
\end{array}
\end{equation}
Here $\kappa_0$ is the asymptotic value of the six-dimensional 
string coupling constant, 
$\kappa_0\sim {g_{str}\over \sqrt{V_{K3}}}$ 
(we will not need the precise expression for it). In the transformation 
law for $\psi$, the vector $H_{+\mu\nu\rho}\in \cwp$ 
acts on $S(\cwp)$ index of $\epsilon$ as gamma-matrices act on spinors.
$D_{\mu}$ is a covariant derivative, which includes the natural connection
on $\cwp$.
In the expression for $\delta\chi$, $\phi$ denotes the point in the
moduli space of scalars. Its derivative $\partial_{\mu}\phi$ acts on 
$\epsilon$ as follows.
There is a natural isomorphism between the tangent bundle 
to $Gr(5,5+21)$ and $\cwm\otimes\cwp$.
Indeed, a tangent vector to $Gr(5,5+21)$
is an infinitesimal rotation $\delta W$ of the plane $W$, 
and it is completely specified by saying 
what is $\delta w\in W^{\perp}$ for each vector $w\in W$. This gives
us a map from $W$ to $W^{\perp}$, or a section of
$\cwm\otimes\cwp^*$. Since $L$ is equipped with a metric, we can
identify $\cwp^*$ with $\cwp$. This means, that we can think of $d\phi$
as an element of $\cwm\otimes\cwp\otimes\Omega^1M$. Then, to act
on $\epsilon\in \Gamma([S_+^{1/2}M\otimes S(\cwp)]_{\bf R})$ by
$\partial_{\mu}\phi\in \cwm\otimes\cwp$, we use the action of $\cwp$ on
$S(\cwp)$ (gamma-matrices act on spinors) and tensor multiplication
by $\cwm$. After acting by the space-time gamma-matrix $\gamma^{\mu}$
we get a section of the bundle 
$(S_-^{1/2}M\otimes S(\cwp))_{\bf R}\otimes\cwm$, which is
where $\chi$ lives.

We want to find a supersymmetric background, corresponding to our
black string. This means that for some  parameters $\epsilon$, 
we should have $\delta_{\epsilon}\psi=0$ and $\delta_{\epsilon}\chi=0$. 
For our solution, there will be eight linearly independent $\epsilon$
satisfying these equations.
We will try the following ansatz for the metric:
\begin{equation}
ds^2=e^{2U(r)}(dt^2-dx^2)-e^{-2U(r)}(dr^2+r^2d\Omega_3^2)
\end{equation}
Let us introduce the vielbein $\{e^t,e^x,e^r,e^1,e^2,e^3\}$,
so that 
$$
ds^2=(e^t)^2-(e^x)^2-(e^r)^2-\sum\limits_{i=1}^3 (e^i)^2
$$
We use the following ansatz for the three-form field strength:
\begin{equation}
\begin{array}{c}
\dss{
H_{ijk}={e^{3U}\over r^3}\epsilon_{ijk}Z,
}
\vspace{5pt}
\\
\dss{
H_{\pm txr}=\pm{e^{3U}\over r^3}Z_{\pm}
}
\end{array}
\end{equation}
The vector $Z$ should be integer-valued, 
$Z\in \Gamma^{5,21}\subset L$, because the fluxes of $H$
through $S^3$ are integers. This vector should be identified with
the charge of the black string.
We look for supersymmetry with the parameter $\epsilon$ 
which depends only on
$r$. Also, we assume that the scalars depend only on $r$. 
First, let us write explicitly the transformations of various components of
$\psi_{\mu}$. 
We will denote $\nabla_{\mu}$ the covariant derivative on $\cwp$.
We do not need to know explicitly what is the spin
connection in $\cwp$, only the space-time spin-connection will be important
for us. We have the following conditions for $\delta\psi=0$
(with $U'={dU\over dr}$):
\begin{equation}\label{variations}
\begin{array}{l}
\dss{
\delta\psi_t={1\over 2}U'e^U\gamma^t\gamma^r\epsilon
+{1\over 2}\kappa_0 {e^{3U}\over r^3}\gamma^x\gamma^r Z_+.\epsilon=0
}
\vspace{5pt}
\\
\dss{
\delta\psi_x=-{1\over 2}U'e^U\gamma^x\gamma^r\epsilon-
 {1\over 2}\kappa_0 {e^{3U}\over r^3}\gamma^t\gamma^r Z_+.\epsilon=0
}
\vspace{5pt}
\\
\dss{
\delta\psi_r=\nabla_r\epsilon+{1\over 2}\kappa_0
{e^{3U}\over r^3}\gamma^t\gamma^x 
Z_+.\epsilon=0
}
\vspace{5pt}
\\
\dss{
\delta\psi_i={1\over 2}U'e^U\gamma^i\gamma^r\epsilon+
{1\over 4}\kappa_0 {e^{3U}\over r^3}
\epsilon_{ijk}\gamma^j\gamma^k Z_+.\epsilon=0
}
\vspace{5pt}
\end{array}
\end{equation}
Let us denote 
$$\hat{Z}_+={Z_+\over \sqrt{||Z_+||^2}}$$
Then, it follows from the first equation in (\ref{variations}), that 
\begin{equation}\label{UandEpsilon}
\begin{array}{c}
\dss{
U'={e^{2U}\over r^3}\sqrt{\kappa_0^2||Z_+||^2},
}
\vspace{5pt}
\\
\dss{
(\hat{Z}_+-\gamma^x\gamma^t)\epsilon=0
}
\end{array}
\end{equation}
The second row in (\ref{variations})
follows from (\ref{UandEpsilon}), and the third gives the equation for
the radial dependence of $\psi$:
\begin{equation}\label{EpsilonFlow}
\nabla_r\psi(r)={1\over 2}\kappa_0{e^{3U}\over r^3}
\psi(r)
\end{equation}
This is consistent with $(\hat{Z}_+-\gamma^x\gamma^t)\epsilon=0$
because for our solution $\hat{Z}_+$ will be covariantly constant
(that is, $\nabla_r\hat{Z}_+(r)=0$).
The last equation in (\ref{variations}) follows from (\ref{UandEpsilon})
if one takes into account that $\epsilon$ is chiral.

Now we turn to the variation of the fermions from tensor multiplets.
We have:
\begin{equation}
\delta\chi=-{1\over 2}e^U\gamma^r(\partial_r\phi).\epsilon -
{e^{3U}\over r^3}\kappa_0 \gamma^1\gamma^2\gamma^3 Z_-\otimes\epsilon=0
\end{equation}
Taking into account the chirality condition 
$\gamma^{123}\epsilon=-\gamma^{txr}\epsilon$ and equations 
(\ref{UandEpsilon}) one can see  that $\delta\chi=0$ if $\phi$
satisfies the following equation:
\begin{equation}\label{ModuliFlow}
d\phi= -2dU {1\over ||Z_+||^2} Z_-\otimes Z_+
\end{equation}
This equation has a very clear geometrical meaning. 
Remember that we represent tangent vectors to the moduli space of scalars 
as linear maps from $W$ to $W^{\perp}$, giving the variations of vectors 
in $W$. 
Suppose that we took a vector $w\in W$ which is orthogonal to $Z$.
Then $(Z_+\cdot w)=0$ and
(\ref{ModuliFlow}) tells us that the variation of $w$ is zero.
Therefore, {\em the subspace $W\cap Z^{\perp}\subset W$ remains constant}. 
We can represent $W=(W\cap Z^{\perp}) \oplus {\bf R}Z_+$. 
When $r$ changes $(W\cap Z^{\perp})$ remains constant
and the one-dimensional subspace ${\bf R}Z_+$ gets rotated
in the plane ${\bf R}Z_+\oplus {\bf R}Z_-$:
\begin{center}
\leavevmode
\epsffile{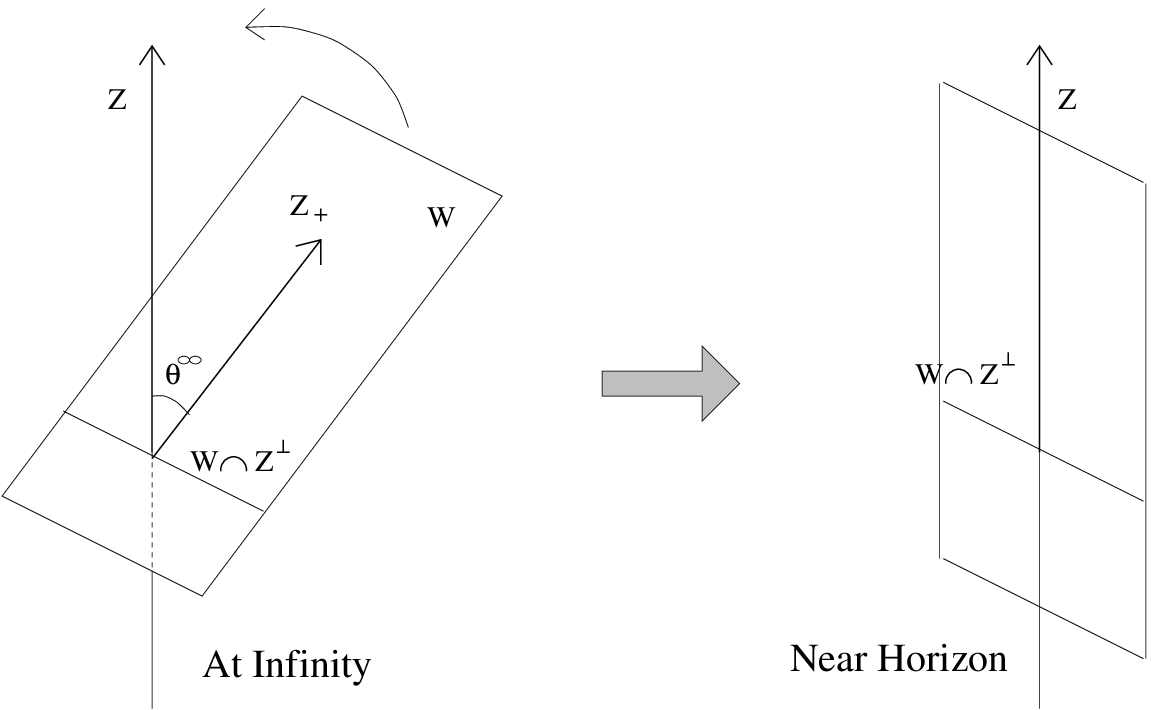}
\end{center}
In particular, 
we see that the near-horizon background is represented by 
$(W\cap Z^{\perp})\oplus {\bf R}Z$.

This is a qualitative picture. Now we want to write the explicit
formula for $\phi(r)$. Let us introduce the hyperbolic
angle $\theta$ between the vector $Z$ and the plane $W$. Let
$\hat{Z}_+=(||Z_+||^2)^{-1/2}Z_+$ and $\hat{Z}_-=(-||Z_-||^2)^{-1/2}Z_-$.
Then,
\begin{equation}\label{DefOfTheta}
\hat{Z}=\cosh\theta \hat{Z}_+ + \sinh\theta \hat{Z}_-
\end{equation}
(this is the definition of $\theta$) and
\begin{equation}
d\hat{Z}_+=-2\tanh\theta dU \hat{Z}_-
\end{equation}
(this is $d\phi$ from (\ref{ModuliFlow}) applied to the vector 
$\hat{Z}_+\in W_+$).
Combining these two equations, we have
\begin{equation}
d\theta=2\tanh\theta dU
\end{equation}
This equation together with the second equation from
(\ref{UandEpsilon}) determines the dependence of $U$ and $\theta$ on $r$:
\begin{equation}\label{AttractorEquations}
\begin{array}{c}
\dss{
U'={e^{2U}\over r^3}\sqrt{\kappa_0^2 ||Z||^2}\cosh\theta,
}\vspace{5pt}
\\
\dss{
\theta'=2{e^{2U}\over r^3}\sqrt{\kappa_0^2 ||Z||^2}\sinh\theta
}
\end{array}
\end{equation}
Adding and subtracting these two equations, we get:
\begin{equation}
{d\over dr}(2U\pm\theta)=
2 \sqrt{\kappa_0^2 ||Z||^2}{e^{2U\pm\theta}\over r^3}
\end{equation}
The result of integration depends on two constants, $c_+$ and $c_-$:
\begin{equation}\label{UandTheta}
\begin{array}{rcl}
\dss{
e^{-2U}}&\dss{=}&\dss{
\left(c_++{\kappa_0\sqrt{||Z||^2}\over r^2}\right)^{1/2}
              \left(c_-+{\kappa_0\sqrt{||Z||^2}\over r^2}\right)^{1/2}
}\vspace{5pt}
\\
\dss{
e^{\theta}}&\dss{=}&\dss{\left(\kappa_0\sqrt{||Z||^2}+c_-r^2\over 
\kappa_0\sqrt{||Z||^2}+c_+r^2\right)^{1/2}
}\vspace{5pt}
\end{array}
\end{equation}
We want $U=0$ at spatial infinity, therefore $c_+c_-=1$. 
The ratio $c_+/c_-$ is related to the asymptotic value of moduli.
This gives:
\begin{equation}
c_{\pm}=e^{\mp\theta_{\infty}}
\end{equation}

{\bf Two Examples.}

1) The system of $Q_0$ $D$-strings
and $Q_4$ $D5$-branes with $B=0$. (The corresponding supergravity 
solution has been found in \cite{CM}). 
The plane $W$ is generated by three vectors orthogonal to $Z$ and the
vector $(1,0,V)$. In this situation only the volume of $K3$ changes
with $r$, the shape remains fixed.
We have:
\begin{equation}
\left({1\over\sqrt{V}},\sqrt{V}\right)=
\cosh\theta \left(\sqrt{Q_4\over Q_0},\sqrt{Q_0\over Q_4}\right)+
\sinh\theta \left(\sqrt{Q_4\over Q_0},-\sqrt{Q_0\over Q_4}\right)
\end{equation}
Therefore, 
\begin{equation}\label{cplus}
e^{-\theta(r)}=\sqrt{Q_4 V(r)\over Q_0},\;\;\;
c_+=\sqrt{Q_4 V^{\infty}\over Q_0}
\end{equation}
Now, (\ref{UandTheta}) gives
\begin{equation}\label{BisZero}
\begin{array}{rcl}
\dss{
e^{-2U(r)}}&\dss{=}&\dss{
\left(1+\sqrt{2V^{\infty}}\kappa_0{Q_4\over r^2}\right)^{1/2}
\left(1+\sqrt{2\over V^{\infty}}\kappa_0{Q_0\over r^2}\right)^{1/2}
}\vspace{5pt}
\\
\dss{
V(r)}&\dss{=}&
\dss{V^{\infty}\;{1+\sqrt{2\over V^{\infty}}\kappa_0{Q_0\over r^2}
\over 1+\sqrt{2V^{\infty}}\kappa_0{Q_4\over r^2}}
}\vspace{5pt}
\end{array}
\end{equation}

2) Although we will not consider adding symmetric branes and
fundamental strings in this paper, we will now give the 
corresponding solution as an example. Consider a system of
$q_0$ fundamental strings and $q_4$ NS5 branes wrapped on $X$.
Using the definition (\ref{DefOfTheta}) of $\theta$ and formula
(\ref{FifthVector}) from Appendix A, we have the equation for 
$\theta(r)$:
\begin{equation}
\left(\kappa(r),{1\over \kappa(r)}\right)
=\cosh\theta(r) \left(\sqrt{q_4\over q_0},\sqrt{q_0\over q_4}\right)+
 \sinh\theta(r) \left(\sqrt{q_4\over q_0},-\sqrt{q_0\over q_4}\right)
\end{equation}
This gives $c_+={1\over\kappa_0}\sqrt{q_4\over q_0}$ and the
following dependence of the metric and the coupling constant
on $r$:
\begin{equation}
\begin{array}{rcl}
\dss{
e^{-2U(r)}}&\dss{=}&\dss{\left(1+{\sqrt{2}q_4\over r^2}\right)^{1/2}
\left(1+\kappa_0^2{\sqrt{2}q_0\over r^2}\right)^{1/2}
}\vspace{5pt}
\\
\dss{
\kappa^2(r)}&\dss{=}&\dss{\kappa_0^2\;{r^2+\sqrt{2}q_4\over
r^2+\sqrt{2} \kappa_0^2 q_0}
}\vspace{5pt}
\end{array}
\end{equation}

{\bf Lagrangean Interpretation.}

The attractor equations have a Lagrangean interpretation. 
Indeed, let us introduce the ``time'' $\tau=-{1\over r^2}$.
The attractor equations (\ref{AttractorEquations}) imply that:
\begin{equation}
\begin{array}{c}
\dss{
{d^2U\over d\tau^2}={1\over 2} 
\kappa_0^2 ||Z||^2 e^{4U}(\cosh^2\theta+\sinh^2\theta),}
\\ \dss{
{d^2\theta\over d\tau^2}=2 e^{4U} \kappa_0^2 ||Z||^2 \sinh\theta\cosh\theta}
\end{array}
\end{equation}
These second order equations may be derived from the following 
Lagrangian:
\begin{equation}
{\cal L}={1\over 4}\left({d\theta\over d\tau}\right)^2+
\left({dU\over d\tau}\right)^2+
{\kappa_0^2||Z||^2\over 4}e^{4U}(\cosh^2\theta+\sinh^2\theta)
\end{equation}
The ``potential energy'' is 
\begin{equation}
E_{pot}=-{1\over 4} \kappa_0^2 e^{4U}(||P_+Z||^2-||P_-Z||^2)
\end{equation}
The expression $||P_+Z||^2-||P_-Z||^2=2||P_+Z||^2-||Z||^2$
is, up to the coefficient and the constant term, the square of the 
tension of the black string. 
At the attracting point, that is when $P_+Z=Z$, this tension
reaches its minimum. Also, we can see from our solution 
(\ref{UandTheta}), that $e^{4U}$ is zero at the horizon.
Therefore, the ``motion'' in $\tau$, corresponding
to the geometry of the black string, may be considered
as ``climbing'' up the potential. The initial velocity should be
adjusted in such a way that getting to the top requires infinite
time. 

In what follows, we will not need the full solution for $U(r)$ and $W(r)$.
The only thing important for us is that $W$ changes with $r$
in such a way that $W(r)\cap Z^{\perp}$ remains constant, and also that
the near-horizon $W$ is $W^h=(W\cap Z^{\perp})\oplus {\bf R}Z$.

\subsection{When can we use the supergravity approximation?}
We want to be able to use supergravity equations  until
we reach small enough values of $r$, where the moduli are close to
their fixed values.  
The low energy effective action does not receive any quantum corrections 
because of the supersymmetry. Therefore, the supergravity approximation 
is valid when  we can neglect higher derivative corrections. 
We will assume that the near-horizon string coupling is weak, therefore
the main source of higher derivative corrections is the string perturbation
theory. These corrections behave like a positive power of $l_{str}\over r$
where $r$ is the typical radius of curvature. Therefore, the condition
that supergravity is applicable is that the size of the AdS region is much
larger then $l_{str}$. Let us explain what we mean by the ``size of the
AdS region''. The metric for the $D1D5$ solution contains factors of
the form
$$
\left(1+\kappa_0{e^{-\theta_{\infty}}\sqrt{||Z||^2}\over r^2}\right)
\left(1+\kappa_0{e^{\theta_{\infty}}\sqrt{||Z||^2}\over r^2}\right)
$$
We are in AdS regime when we can approximate this expression
as ${\kappa_0^2||Z||^2\over r^4}$. Assuming $\theta_{\infty}>0$,
we see that the AdS regime starts at
\begin{equation}
r^2<r_{cr}^2=\kappa_0 e^{-\theta_{\infty}}\sqrt{||Z||^2}
\end{equation}
The condition for supergravity to be valid is that $r_{cr}>>l_{str}$.
Now let us estimate $l_{str}$ in the near-horizon region. 
It is convenient to start with estimating the size of the core 
of the $D$-string. 
For the typical attractor, all the moduli of the near-horizon torus
are of the order one, therefore all the black $D$-strings (for example,
$D1$, $D3$ wrapped on a two-cycle and $D_5$ wrapped on ${\bf T}^4$) have
approximately the same string length and the same size of the core. 
For the typical D-string in the typical background the size of the core is 
\begin{equation}\label{DvsF}
r_{core, \; D}^2 \sim g_{str} l_{str}^2
\end{equation}
As a typical D-string, let us choose the string with the charge 
${Z\over \sqrt{||Z||^2}}$. This charge  is usually not integer, 
but we need only the supergravity solution. It has the metric with the 
characteristic factor
\begin{equation}
\left(1+\kappa_0 e^{-\theta_{\infty}}
\left[{\sqrt{||Z||^2}\over \vec{r}^2}+{1\over (\vec{r}-\vec{r}_0)^2}
\right]
\right)
\left(1+\kappa_0e^{\theta_{\infty}}
\left[{\sqrt{||Z||^2}\over \vec{r}^2}+{1\over (\vec{r}-\vec{r}_0)^2}
\right]
\right)
\end{equation}
One can see from this expression that the size of the core for the
D-string at the coordinate distance $r_0$ from the origin is of the order 
\begin{equation}
r_{core,\;D}^2\sim r_0^2 /\sqrt{||Z||^2}
\end{equation}
This equation together with (\ref{DvsF}) implies
\begin{equation}
l_{str}^2(r_0)\sim {r_0^2\over \sqrt{||Z||^2} g_{str}(r_0)}
\end{equation}
Since $\kappa=\mbox{const}$ for the D-string, we have 
\begin{equation}
g_{str}(r_0)=\sqrt{V(r_0)\over V^{\infty}} g_{str}^{\infty}\sim 
{g_{str}^{\infty}\over \sqrt{V^{\infty}}}
\end{equation}
This implies that for $r_0\sim r_{cr}$ we have:
\begin{equation}\label{Estimate}
{r_{cr}^2\over l_{str}^2(r_{cr})} \sim 
{g_{str}^{\infty}\sqrt{||Z||^2}\over \sqrt{V^{\infty}}}
\end{equation}
(Remember that we measure all the lengths for the torus in string units).
The right hand side of this expression has a clear physical meaning.
It is the t'Hooft coupling constant for the $1+5$-dimensional 
Super Yang-Mills on ${\bf R}^{1,1}\times \widetilde{\tfour}$ on
the scale of the size of $\widetilde{\tfour}$. (Here $\widetilde{\tfour}$
is the noncommutative torus ``dual'' to the torus $\tfour$ at infinity
as explained in Section 5).
Our estimate (\ref{Estimate}) tells us that in the regime
when we can trust the supergravity solution in the AdS region, the 
six-dimensional noncommutative Yang-Mills describing the $D1D5$ system
Moduli pois necessarily strongly coupled on the compactification scale. 
Therefore, we cannot trust the classical dimensional reduction.
We conjecture that the shape of the target
space of the sigma-model on ${\bf R}^{1,1}$ which is the dimensional
reduction of this six-dimensional theory  actually does not depend on 
the six-dimensional coupling constant. We will give some evidence 
confirming this conjecture in Section 6. 

\section{Argument based on supersymmetry.}
In this section we will prove using  supersymmetry  that 
two backgrounds flowing to the same attractor give the same 
low energy effective theory on the worldvolume of the black string.
Our argument is very close to the one used
in \cite{MaldacenaLE} to prove independence of the hypermultiplet metric
on the scalars in the vector multiplet.

We will study some auxiliary brane configuration.
It consists of the system of $N$ parallel black strings, all having the
same charge $Q$. This system has a Coulomb branch. Let us consider
the submanifold of this Coulomb branch, where $N-1$ black strings
sit at one point (the origin), and one is moving around:
\begin{center}
\leavevmode
\epsffile{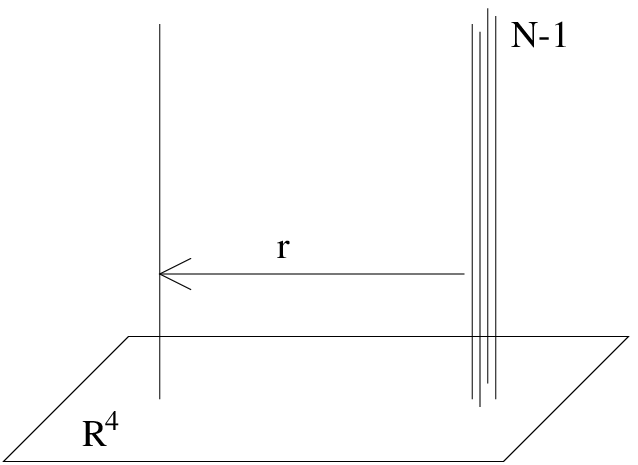}
\end{center}
We will call that black string which is moving around the ``single
black string''. We want to look at the
worldsheet theory of this single black string. 
The low-energy worldsheet theory is some sigma model. Consider the
target space $\trgt$ of this sigma-model. It can be parametrized
by the position in the transversal ${\bf R}^4$ plus the internal 
degrees of freedom, which specify the instanton on ${\bf T}^4$.
Notice that the size of the moduli space of instantons on ${\bf T}^4$
is of the order ${\sqrt{V}\over g_{str}}$ and does not depend on
$N$, while the characteristic distances in the transversal ${\bf R}^4$
are growing with $N$. (From our solution, we see that 
$R^2\sim N\sqrt{||Z||^2}$.) Therefore, in the limit $N\to\infty$, we can
consider the internal degrees of freedom as ``fast'', and the motion
along ${\bf R}^4$ as ``slow''. Moreover, in this limit the
single black string sitting at some point $x\in{\bf R}^4$ 
does not feel that the background changes with the distance from
the origin. Therefore,  the configuration space 
of the internal degrees of freedom
should be approximately the same Hyper-K\"ahler manifold 
as for the black string in flat space. 
In other words, in the limit $N\to\infty$ we can think of $\trgt$ 
as a bundle with the base ${\bf R}^4$ and the fiber the moduli space 
${\cal M}(x)$ of 
internal degrees of freedom of $D1D5$ wrapped on ${\bf T}^4(x)$:
\begin{equation}\label{bundle}
\widehat{\cal M}\stackrel{\cal M}{\rightarrow} {\bf R}^4
\end{equation}
The dependence of ${\bf T}^4(x)$ on the point $x\in {\bf R}^4$ 
(and therefore the dependence of ${\cal M}(x)$ on $x$) is dictated
by the supergravity solution for $N-1$ black strings sitting at the origin.

Our arguments will go as follows.
We will prove, using supersymmetry, that although the moduli of the
torus ${\bf T}^4$ depend on the distance from the origin, the fiber
${\cal M}$ does not. But in the limit $N\to\infty$, the fiber
${\cal M}(x_0)$ over the given point $x_0\in {\bf R}^4$
is the same as the target space ${\cal M}_0$ for the
$D1D5$ in the flat background with the torus ${\bf T}^4 (x)\equiv
{\bf T}^4 (x_0)$. Therefore, independence of ${\cal M}(x_0)$ on $x_0$
implies that the target space for $D1D5$ is the same for all the
backgrounds ${\bf T}^4={\bf T}^4(x_0)$, $x_0\in {\bf R}^4$. 
Since ${\bf T}^4(x_0)$ depends only on the distance $r$ of $x_0$ from the
origin, we had proven that the moduli space of $D1D5$ is the same
for the backgrounds from the given one-parametric family ${\bf T}^4(r)$,
$r\in [0,\infty]$. In particular, it is the same for ${\bf T}^4(r=0)$
and ${\bf T}^4(r=\infty)$. This implies, that all the backgrounds at 
infinity $r=\infty$
which flow to the same near-horizon ${\bf T}^4(r=0)$ give the
same moduli space. 

Let us proceed with the proof.
We want to understand what kind of restrictions the supersymmetry imposes
on the geometry of our target space $\trgt$. It is useful to start with
considering the ``flat'' sigma-model, which describes the dynamics
of $D1D5$ in flat space (without $N-1$ strings at the origin).
In this case, we would have:
$$
\trgt^0={\bf R}^4\times {\cal M}_0
$$
where ${\cal M}_0$ does not depend on the point of the base. 
Both ${\bf R}^4$ and ${\cal M}_0$ are Hyper-K\"ahler, and therefore
we get the system with $(4,4)$ supersymmetry. In fact, there are two
Hyper-K\"ahler structures on ${\bf R}^4$: one with the self-dual 
Hyper-K\"ahler
forms, and the other with anti-self-dual. Closer examination of the
supersymmetry transformations shows that they are different in the
left- and right-moving sectors: the left-moving $(4,0)$ supersymmetry
uses self-dual complex structures on ${\bf R}^4$, and the right-moving
one uses anti-self-dual complex structures.  

When we introduce $N-1$ black strings at the origin the flat target
space gets corrected; for example, the metric on ${\bf R}^4$ is not
flat anymore. Nevertheless,
our manifold $\trgt$ should still 
support a sigma-model with $(4,4)$ supersymmetry. This follows from the 
fact that our single black string preserves the same supersymmetry as
those $N-1$ black strings which we have added. Therefore, introducing these
$N-1$ black strings does not break any more supersymmetry.
The $(4,4)$ supersymmetry implies the existence of six complex structures, 
three for the left-moving sector, and three for the 
right-moving \cite{GHR}. 
The sigma-model action can include torsion, which is an antisymmetric
tensor $b$ with $db\neq 0$. The lagrangian is:
\begin{equation}
\begin{array}{c}
L=g_{\alpha\beta}(dX^{\alpha}\cdot dX^{\beta})+
2g_{\alpha a}(dX^{\alpha}\cdot dX^a)+
g_{ab}(d\Phi^a \cdot d\Phi^b)+\\
+b_{\alpha\beta}dX^{\alpha}\wedge dX^{\beta}+
2b_{\alpha a}dX^{\alpha} \wedge d\Phi^a +
 b_{ab} d\Phi^a\wedge d\Phi^b
\end{array}
\end{equation}
Here we denote $X^{\alpha}$ the coordinates in ${\bf R}^4$, and $\Phi^a$ the
coordinates in the fiber (the moduli space of instantons).
The complex structures should be covariantly constant, but with respect
to the modified connection, with the torsion $T=db$ added. The sign of
the torsion is opposite for the left and the right connections.
We will write this condition explicitly for the corresponding
K\"ahler forms:
\begin{equation}
\nabla_{\lambda}\omega^{I\pm}_{\mu\nu}\pm 
T_{\lambda\mu}^{\sigma}\omega^{I\pm}_{\sigma\nu}\pm
T_{\lambda\nu}^{\sigma}\omega^{I\pm}_{\mu\sigma}=0
\end{equation}
(the index $I=1,2,3$ distinguishes between the three 
different K\"ahler forms,
and the sign $\pm$ distinguishes between the right and the left sectors).
In our situation, we know something about these six complex structures
from comparison to the ``flat'' case. We know that in the vicinity of a given
fiber, the K\"ahler forms are:
\begin{equation}
\omega^{I\pm}=\omega_{(0)\alpha\beta}^{I\pm}dX^{\alpha}\wedge dX^{\beta}+
\omega_{(0)ab}^{I\pm} d\Phi^a\wedge d\Phi^b+\mbox{small corrections}
\end{equation}
Here $\omega_{(0)\alpha\beta}^{I\pm}$ are the basic self-dual (with plus)
or anti-self-dual (with minus) forms on ${\bf R}^4$, and 
$\omega^{I\pm}_{(0)ab}$ are the K\"ahler forms on flat ${\cal M}_0$.
The corrections are small, because locally our background is almost 
flat\footnote{
The left K\"ahler forms on ${\cal M}_0$ coincide
with the right K\"ahler forms, therefore it might seem strange that
we are using the superscript $\pm$ to distinguish between
$\omega_{(0)ab}^{I+}$ and $\omega_{(0)ab}^{I-}$. 
The reason why we distinguish them is the following. 
The forms $\omega^{I\pm}$ should be covariantly constant (with modified
connection). In particular, they should be covariantly constant along the
radial direction in ${\bf R}^4$. Although the three-dimensional space
generated by $\omega^{I+}_{(0)ab}$ and $\omega^{I-}_{(0)ab}$ coincide,
we do not apriori exclude the possibility that the covariantly constant
bases in these spaces are different in left and write sectors.}.

We also have some information about the torsion. First of all, the
torsion is proportional to the background Ramond-Ramond fields. Indeed,
the corresponding term in the black string effective action violates
parity. In the brane picture, parity turns a black string into an anti-black
string. It is the same as changing the sign of the Ramond-Ramond fields.
Therefore, the torsion should be zero if the Ramond-Ramond fields are
turned off. Also, we know that in flat case the torsion is zero
\footnote{Indeed,  the components of the torsion along ${\bf R}^4$ 
are not allowed
in flat case because of $SO(4)$ invariance. The components of the torsion
along ${\cal M}_0$, $T_{abc}$, are also zero for the following reason.
In the flat case, the only Ramond-Ramond fields are the constant
fields along the torus. We know how they couple to our $D1D5$ system
from supergravity. They couple to the topological numbers. Therefore,
the corresponding contribution to the black string effective action cannot
change under the small deformations of the fields. In other words,
these constant Ramond-Ramond fields generate theta-angles in our
sigma-model, but not a torsion.}. Now we want to see what happens to the
torsion when we put $N-1$ black strings at the origin in ${\bf R}^4$.
First of all, the components of the torsion along ${\bf R}^4$ become
nonzero. For example, $D5$ branes from the pile at the origin
create $B^{RR}$, which couples to $D1$ from the single black string.
Closer examination shows that 
$
T_{ijk}\sim N||Z||^2 \epsilon_{ijk}
$
and the other components in ${\bf R}^4$ are zero. 
Also, we should admit the possibility that the component 
$T_{abr}$ is not zero any more. 
Indeed, the Ramond-Ramond fields along ${\bf T}^4$ change when
we move in the radial direction. This implies  that theta-angles on
${\cal M}$ may change when we move in ${\bf R}^4$. This precisely
means that $T_{abr}\neq 0$. On the other hand, we expect that
$T_{abc}$ is still zero, or at least very small when $N\to\infty$.
Ideed, we have seen that $T_{abc}$ is zero when we put our black
string in flat background. The non-flatness of the background can be 
locally described by the gradient of the background fields, which
we can schematically denote $\partial_{\alpha} \theta$ ($\theta$ denotes
the background fields). From the rotaional symmetry, and assuming
the analytic dependence of the torsion on the background fields, we
estimate:
\begin{equation}
T_{abc}\sim g^{\alpha\beta}\partial_{\alpha}\theta\partial_{\beta}\theta
\end{equation}
In other words, it is of the second order in the gradient of the
background fields. Therefore we will neglect it. 
Also, the other components of the torsion (for example,
$T^i_{ja}$, where $i$ and $j$ are indices from the tangent space
to $S^3$) are zero because of the rotational symmetry of ${\bf R}^4$.

Now we want to consider the variation of the period map (the integrals of
the K\"ahler forms over the two-cycles in ${\cal M}$), when we move
in the radial direction. Let us consider two points, one at the
distance $r_1$ from the origin, and the other at the distance $r_2$
(both on the same line with the origin). Consider the cycle
$c(r_1)$ in ${\cal M}$ at the first point, and the same cycle 
$c(r_2)$ in ${\cal M}$ at the second point. We have, according to the
Stokes theorem,
\begin{equation}
\begin{array}{c}
\int_{c(r_2)} \omega^{I\pm}-\int_{c(r_1)}\omega^{I\pm}=
\pm \int_{c\times I_{[r_1,r_2]}} dx^{\lambda}\wedge dx^{\mu} 
\wedge dx^{\nu} T^{\rho}_{[\lambda\mu}\omega^{I\pm}_{\nu]\rho}
\end{array}
\end{equation}
The leading term on the right hand side is 
\begin{equation}\label{ContributionOfTorsion}
\pm\int dr \int_{c(r)} d\Phi^a\wedge d\Phi^b 
T^c_{r[a}\omega^{I\pm}_{b]c}
\end{equation}
We know that the space of periods over the fiber of $\omega^{I+}$ 
coincides with that of $\omega^{I-}$, modulo small corrections. 
On the other hand, we see from (\ref{ContributionOfTorsion}) that the
change in the periods of $\omega^{I+}$ and the periods of
$\omega^{I-}$ as we move along $r$ is opposite. The only possibility
to reconcile these two observations is that $T^c_{r[a}\omega^{I\pm}_{b]c}$
is either zero or some combination $A^I_J\omega^{J\pm}$. 
In any case, {\em it follows that the space of periods does not
depend on $r$}. 

The condition
\begin{equation}
T^c_{r[a}\omega^{I\pm}_{b]c}=\mbox{linear combination of}\;\;\omega^{J\pm}
\end{equation}
is a restriction on the torsion. This restriction implies that the 
Hyper-K\"ahler structure
of the fiber does not depend on $r$. 
It is automatically satisfied
when $T^c_{ra}$ is zero --- in other words, when the theta-angles on
${\cal M}(x)$ do not depend on the position in $x\in{\bf R}^4$.
We believe that this is what actually happens:
\begin{equation}
T_{ra}^b=0
\end{equation}
Indeed, our analysis
of supergravity solution shows that of the 
eight Ramond-Ramond fields on the
torus, there are seven linear combinations which do not depend on $r$.
It is natural that the seven worldsheet theta-angles depend precisely on 
these seven combinations of the Ramond-Ramond fields. We will check it 
in the case when the size of the torus is very small at the end of
Section 5, using the methods of noncommutative geometry.

\section{Near horizon geometry.}
In this section we will study the correspondence between the
background fields far away from the black string (we will call them
``asymptotic background'') and the background
fields at the horizon. The answer is given by the formulas
(\ref{NHvsAs}) and (\ref{AsvsNH}).
We will explain under which conditions two different
asymptotic backgrounds flow to the same near-horizon background
(equations (\ref{WhichAreEquivalent})).
We will show that for each background there exists a background
with the small size torus which flows to the same near-horizon geometry.
To simplify the discussion, we will first turn off the Ramond-Ramond fluxes
and study only the ``perturbative'' moduli, parametrized by 
$Gr(4,4+20)$. We will include the Ramond-Ramond fields at the end of this
section. The correspondence between the Ramond-Ramond fluxes at infinity
and at the horizon is given by (\ref{CiH})

\subsection{Moduli of string perturbation theory.}
We will use the correspondence between the string theory backgrounds and
the planes $W$, which was first discussed in \cite{Aspinwall}
and then made more precise in \cite{DanAndSanjaye} and \cite{Dijkgraaf}.
The 
plane $W$ is generated by the vectors $\vec{v}$ and $v^0$, given by:
\begin{equation}\label{CFTModuli}
\begin{array}{c}
\dss{
\vec{v}=(0,\vec{\omega},-B\cdot\vec{\omega}),}
\\
\dss{v^0=(1,B,V-{1\over 2} B\cdot B)}
\end{array}
\end{equation}
Here $\vec{\omega}=(\omega^1,\omega^2,\omega^3)$ are the three 
K\"ahler forms,
$V$ is the volume and $B$ is the $B$ field. 
Now suppose that we are given the charge vector $Z$:
\begin{equation}
Z=(Q_4,Q_2,Q_0)
\end{equation}
According to the previous section, the near-horizon geometry corresponds 
to the plane $W^h=(W\cap Z^{\perp})\oplus {\bf R} Z$. 
Let us describe this $W^h$ explicitly.

We begin with introducing a useful notation.
The space $\Lambda_+^2{\bf T}^4$ of self-dual forms on our torus 
is generated by $\omega^i$. Let us consider the two-dimensional subspace
\begin{equation}\label{LC}
\lc=\{ \omega\in\Lambda_+^2{\bf T}^4 | ((Q_2-Q_4 B)\cdot\omega)=0 \}
\end{equation}
We have denoted this space $\lc$ for the following reason.
As explained in \cite{Aspinwall}, the choice of the two-dimensional 
subspace in $H^2(X, {\bf R})$ (where $X$ is $K3$ or $\tfour$) 
is equivalent to the choice of the complex structure up to a sign
(to fix a sign, we have to orient the two-plane).
Therefore, the charge vector $Z$ gives our torus a preferred complex
structure modulo sign, $\lc$.
The orthogonal space to $\Lambda_{\bf C}$ in 
$\Lambda^2_+X$ is one-dimensional. Therefore, there are two ways
to choose the K\"ahler form, which differ by a sign. The choice of the
sign of the K\"ahler form also fixes the sign of the complex structure.
(In the case of torus this may be understood as follows: we need the
value of the K\"ahler form on any tangent bivector to 
any holomorphic surface to be positive).
Our preferred complex structure has a clear geometrical meaning in the case
when $B=0$. In this case, the bound state of $D1D5$ may be described
as an instanton configuration on $U(Q_4)$ gauge theory with the
topological charges $(Q_2,Q_0)$. The corresponding field strength
is of the type $(1,1)$ in the complex structure
selected by the condition $(Q_2\cdot\omegc)=0$. Our definition
(\ref{LC}) is the generalization of this complex structure for $B\neq 0$.

Now we describe $W\cap Z^{\perp}$. This space contains a two-dimensional
subspace
\begin{equation}
W_{\bf C}=\{ (0,\omega,-(B\cdot\omega)) | \omega\in \lc \}
\end{equation}
Since $W\cap Z^{\perp}$ is three-dimensional, we need one more vector.
Choose $\omegr\in\Lambda^2_+\tfour$, which satisfies two conditions:
\begin{equation}
\begin{array}{cl}
\dss{1)}& \dss{\omegr\perp\lc} 
\vspace{5pt}\\
\dss{2)}& ||\omegr||^2=1
\end{array}
\end{equation}
The two-form $\sqrt{V}\omegr$ is the K\"ahler form up to sign.
The vector
\begin{equation}\label{vr}
v'_{\bf R}=(1,B+s\omegr,V-{1\over 2}B\cdot B - s B\cdot\omegr)
\end{equation}
where
\begin{equation}\label{s}
s={1\over 2Q_4}
\frac{||Q_2-Q_4B||^2-||Z||^2-2Q_4^2V}{((Q_2-Q_4B)\cdot\omegr)}
\end{equation}
belongs to $W\cap Z^{\perp}$. Moreover, it is orthogonal to
$\wc$. 
Therefore, $W\cap Z^{\perp}$ is generated by the two-dimensional
space $\wc$ and the vector $v'_{\bf R}$.
The near-horizon $W^h$ is generated by this $W\cap Z^{\perp}$ and
$Z$. 
Let us introduce the following linear combinations of $v'_{\bf R}$ and $Z$:
\begin{equation}
\begin{array}{l}
\dss{v_{\bf R}^h={Q_4 v_{\bf R}'-Z\over 
\sqrt{||Z||^2+ Q_4^2(2V+s^2)}}}
\vspace{5pt}
\\
\dss{v_0^h={||Z||^2 v_{\bf R}' + Q_4 (2V+s^2) Z \over
       ||Z||^2 + Q_4^2 (2V+s^2)}}
\end{array}
\end{equation}
They are useful, because they are orthogonal to each other and $\wc$, 
$v_{\bf R}^h$ is of the form $(0,*,*)$ and $||v_{\bf R}||^2=1$,
and $v_0^h$ is of the form $(1,*,*)$. It follows from the relation
(\ref{CFTModuli}) between the background fields and the plane $W$
that the near-horizon K\"ahler form,
$B$ field and volume are related to the components of $v_{\bf R}^h$ and
$v_0^h$ in the following way:
\begin{equation}
\begin{array}{l}
\dss{v_{\bf R}^h=(0,\omegr^h, *)}\vspace{5pt}\\
\dss{v_0^h=(1, B^h, *)}\vspace{5pt}\\
\dss{||v_0^h||^2=2V^h}\vspace{5pt}
\end{array}
\end{equation}

{\bf The near-horizon background as a function 
of the background at infinity.} 

This gives the following expressions for the near-horizon background fields
in terms of the asymptotic background fields:
\begin{equation}\label{NHvsAs}
\begin{array}{l}
\dss{\lc^h=\lc}
\vspace{5pt}\\
\dss{\omegr^h={Q_4(B+s\omegr)-Q_2\over \sqrt{||Z||^2+Q_4^2(2V+s^2)}}}
\vspace{5pt}\\
\dss{B^h={||Z||^2(B+s\omegr)+(2V+s^2)Q_4Q_2\over
     ||Z||^2+Q_4^2(2V+s^2)}}
\vspace{5pt}\\
\dss{2V^h={||Z||^2(2V+s^2)\over ||Z||^2+Q_4^2(2V+s^2)}}
\vspace{5pt}
\end{array}
\end{equation}
The sign of the square root is not fixed (as we have explained before,
the choice of the sign of the K\"ahler form corresponds to the choice
of the sign of the complex structure).
One can see that the near-horizon background fields satisfy the
attractor conditions:
\begin{equation}\label{AttractorFields}
Q_4B^h-Q_2=\sqrt{||Z||^2-2Q_4^2V^h}\omegr^h
\end{equation}

{\bf Which backgrounds at infinity flow 
to the given near-horizon background?}

Let us now invert (\ref{NHvsAs}) and
find the  backgrounds flowing to the given attractor.
The condition (\ref{AttractorFields}) tells us that 
the near-horizon $B$-field is expressed
in terms of the near-horizon volume and K\"ahler form. Therefore, it
is natural to parametrize attractors by  $(V^h,\omegr^h,\lc^h)$.
There is a twenty-parameter family of backgrounds 
 flowing to $(V^h,\omegr^h,\lc^h)$. They may be
parametrized by a real number $s$ and the K\"ahler form $\omegr$.
The $B$-field, the volume and the complex structure are determined
from the following equations: 
\begin{equation}\label{AsvsNH}
\begin{array}{rcl}
\dss{\lc}&\dss{=}&\dss{\lc^h}
\vspace{5pt}\\
\dss{B+s\omegr}&
\dss{=}&
\dss{{Q_2\over Q_4}+{||Z||^2\over Q_4\sqrt{||Z||^2-2Q_4^2V^h}}\omegr^h}
\vspace{5pt}\\
\dss{V+{s^2\over 2}}&\dss{=}&\dss{{||Z||^2\over ||Z||^2-2Q_4^2V^h}V^h}
\vspace{5pt}
\end{array}
\end{equation}
From the last of these relations we get a restriction on $V$:
\begin{equation}
V\leq {||Z||^2\over ||Z||^2-2Q_4^2V^h}V^h
\end{equation}
Therefore, for any given attractor there is an upper limit on the
volumes of the possible tori at infinity.

{\bf When are two backgrounds at infinity equivalent?}

We will call two backgrounds  equivalent if 
they flow to the same background at the horizon.  
Our equivalence classes are 20-dimensional subspaces in 
$Gr(4,4+20)$ in the case of $K3$, or 4-dimensional subspaces
in $Gr(4,4+4)$ in the case of $\tfour$. 
Indeed, given a space $W\cap Z^{\perp}$, we need
to specify a single line in ${\bf R}^{4,20}/(W\cap Z^{\perp})$,
in order to specify $W$. Therefore, the equivalence class 
is itself a Grassmanian manifold, $Gr(1,1+20)$.  
The space of equivalence classes is also a Grassmanian, $Gr(3,3+20)$. 
It follows from (\ref{NHvsAs}) or (\ref{AsvsNH}) that
for the background with parameters $B',\omega'^i$ 
to be equivalent to the background $B,\omega$, the following conditions
should be satisfied:
\begin{equation}\label{WhichAreEquivalent}
\begin{array}{cl}
\dss{1)} & \dss{\mbox{Complex structures are the same},}
\vspace{5pt} \\
\dss{2)} & \dss{B'+s'\omegr'=B+s\omegr,}
\vspace{4pt}\\
\dss{3)} & \dss{V'+{(s')^2\over 2}=V+{s^2\over 2}}
\vspace{5pt}
\end{array}
\end{equation}
(Apriori, the complex structures are the same modulo sign. But since
the equivalence classes are connected, it is natural to choose 
the sign to be the same.)
The third condition follows from the first two and the definition
(\ref{s}) of $s$, therefore we have in total $40+20=60$ conditions
for $X=K3$, or $8+4=12$ for ${\bf T}^4$. (This is the dimension
of $Gr(3, 3+20)$ or $Gr(3,3+4)$, respectively.) 

{\bf Two Examples.}

1) As a consistency check, let us see what happens when $V>>1$. 
In this case, the near-horizon plane $W^h$ is generated by the vectors:
\begin{equation}
v^h_{\bf C}=(0,\omega_{\bf C},0),\;\; v^h_{\bf R}\simeq
(0,\omega_{\bf R},0),\;\; v_0^h\simeq (1,Q_2/Q_4,Q_0/Q_4)
\end{equation}
This plane does not depend on $B$. Therefore, for the large volume torus
the moduli space of instantons does not depend on $B$. 
This is what we expect. Indeed, in this limit we expect
that the system is described by the modified six-dimensional
Yang-Mills action:
\begin{equation}
S_B=\mbox{const}+ {1\over \alpha' g_{str}}\int \mbox{tr}\, (F-B)^2 d^6x
\end{equation}
The corresponding equations of motion do not depend on $B$,
and the moduli space of solutions is the same as for $B=0$.

2) Now consider the equivalence of backgrounds with
very small torus, $V=V'=0$. Then, we have simply $s=s'$. 
This allows us to rewrite the equivalence conditions in somewhat 
simpler form. Namely, the two small tori are equivalent if and only if:
\begin{equation}\label{SmallVEquivalent}
\begin{array}{cl}
\dss{1)} & \dss{\mbox{They have the same complex structure}}
\vspace{5pt}
\\
\dss{2)} & \dss{\mbox{They have the same two-form}
\;\;\;\;\omegr-2{(\omegr\cdot(Q_2-Q_4B))\over 
||Q_2-Q_4B||^2-||Z||^2}(Q_2-Q_4B)}
\end{array}
\end{equation}

{\bf Each equivalence class contains small tori.}

Given the torus of the finite size, it is easy to find the corresponding
family of small tori. Let us denote the background fields for the finite
size torus $(B,\omegr,\lc)$, and the background fields for the
small tori $(B^0,\omegr^0,\lc^0)$. 
From the last of equations (\ref{WhichAreEquivalent}),
we express $s^0$ for the family of small tori in terms of the volume 
and $s$ for  our finite size torus:
\begin{equation}
s^0=\sqrt{2V+s^2}
\end{equation}
Our family of small tori can be parametrized by $\omegr^0$, which
is constrained to be orthogonal to $\lc^0=\lc$.
The field $B^0$ is found from the second equation 
in (\ref{WhichAreEquivalent}) which tells us that $B^0+s^0\omegr^0$ 
is equal to $B+s\omegr$.
(If $s^0$ is related to $s$ by the last of eqs. 
(\ref{WhichAreEquivalent}) and $(B^0,\omegr^0)$ satisfy the second of 
(\ref{WhichAreEquivalent}), and $s$ is related to $(\omegr, B, V)$ by 
the formula (\ref{s}),
then $s^0$ is also related to $(\omegr^0, B^0, V^0=0)$ by the same formula
(\ref{s})).
We see that, indeed, there are small tori in each equaivalence class.
This allows us
to describe the dynamics of branes wrapped on a finite torus in terms of
branes wrapped on a small torus. The latter is related to the 
noncommutative geometry, as reviewed in the next section.

\subsection{Turning on the Ramond-Ramond fields.}
Now we want to include the Ramond-Ramond fluxes.
We will use the notation for the vectors in ${\bf R}^{5,21}$ which is 
explained at the end of Appendix A. 
Notice that turning on the Ramond-Ramond fields does not
change the near-horizon values of the perturbative moduli, calculated
in the previous subsection. Indeed, the four-plane specifying
the perturbative moduli is given by 
$(W\cap v^{\perp})/v$ where $v={[}(0,0,0),(0,1){]}$. But 
$v\in Z^{\perp}$, which means that the reduction to $v^{\perp}/v$
commutes with the operation $W\to (W\cap Z^{\perp})\oplus {\bf R}Z$
used to compute the near-horizon moduli.
We want to answer the question: what are the near-horizon RR fluxes,
given the RR fluxes at infinity? Consider the plane $W^{\infty}$ 
corresponding to the background at infinity. It is generated
by five vectors the fifth of which is:
\begin{equation}
v^5=[(-C_0,C_2,C_4),(1,\kappa^{-2}-{1\over 2}||{\cal C}||^2)]
\end{equation}
Let us denote $W_4=W\cap v^{\perp}$, $v={[}(0,0,0),(0,1){]}$.
In particular $W_4^{\infty}$ is the plane generated by the first 
four vectors $v^0,\ldots,v^3$.  Consider the vector
\begin{equation}\label{VFiveH}
v^5_h=v^5-{(v^5\cdot Z)\over ||P_{W^{\infty}_4}Z||^2} P_{W^{\infty}_4}Z
\end{equation}
We claim that $v^5_h=[(-C^h_0,C^h_2,C^h_4),
(1,\kappa_h^{-2}-{1\over 2}||{\cal C}||^2)]$, 
where $C^h_i$ and $\kappa_h$ are the near-horizon values of
the Ramond-Ramond fluxes and the six-dimensional string coupling constant.
Let us prove it. The subspace $W^{\infty}_4\subset W^{\infty}$ 
consists of the vectors
of the form: $[(*,*,*),(0,*)]$. We can characterize $v^5$ as the vector
of the form $[(*,*,*),(1,*)]$, with the conditions $v^5\in W$
and $v^5\perp W_4$. The vector $v^5_h$ defined in (\ref{VFiveH})
belongs to $W^{\infty}\cap Z^{\perp}$. 
Since $W\cap Z^{\perp}$ is constant, $v_5$ 
also belongs to the near-horizon five-plane $W^h$. 
Because of the attractor condition, $W^h$ contains $Z$. 
Since we do not consider fivebranes and fundamental strings, the last
two components of $Z$ are zero. Therefore $Z$ belongs to the 
near-horizon four-plane, $Z\in W^h_4$. Also, notice that 
$W_4^{\infty}\cap Z^{\perp}$ is three-dimensional. Now it follows that
the near-horizon $W_4^h$ is generated by $Z$ and 
$W_4^{\infty}\cap Z^{\perp}$. Besides being orthogonal to $Z$, our vector
$v^5_h$ is also orthogonal to $W_4^{\infty}\cap Z^{\perp}$ (indeed,
$v^5$ is orthogonal to $W_4^{\infty}$ and $P_{W_4^{\infty}}Z$ is
orthogonal to $W_4^{\infty}\cap Z^{\perp}$).
Therefore, $v^5_h$ is orthogonal to
$W_4^h$. Also, it is clearly of the form $[(*,*,*),(1,*)]$.
Because of our characterization of $v^5$, these properties imply that
the vector $v^5_h$ defined in (\ref{VFiveH}) is, indeed, the near-horizon
$v^5$.

Straightforward computation using (\ref{VFiveH}) gives (when $V=0$):
\begin{equation}\label{CiH}
\begin{array}{l}
\dss{C_0^h=C_0+t,}\vspace{5pt}\\
\dss{C_2^h=C_2-tB,}\vspace{5pt}\\
\dss{C_4^h=C_4+{t\over 2}||B||^2}\vspace{5pt}
\end{array}
\end{equation}
where we have denoted:
\begin{equation}\label{WhatIsTea}
t={-Q_0C_0+Q_4C_4+(Q_2\cdot C_2)\over
   Q_0+(Q_2\cdot B)-{1\over 2}Q_4||B||^2}
\end{equation}
One can check that the total flux of the near-horizon RR fields through 
our branes is zero:
\begin{equation}
-Q_0C_0^h+Q_4C_4^h+(Q_2\cdot C_2^h)=0
\end{equation}
This is the attractor condition for the Ramond-Ramond fields.
The relation (\ref{CiH}) between the asymptotic and the near-horizon 
values of the Ramond-Ramond fields may be rewritten in terms of 
$C=C_0+C_2+C_4$:
\begin{equation}
C^h=C+te^{-B}
\end{equation}
Notice that the expressions (\ref{CiH}) for $C_i^h$ do not involve
$\omegr$ (when $V=0$). 
It has been shown in \cite{Dijkgraaf} that the near-horizon
values of the Ramond-Ramond fluxes correspond to the fluxes of the
$B$-fields in the sigma-model on the moduli space of instantons.
In other words, they are the theta-angles of the instanton sigma-model. 
When $B=0$, (\ref{CiH}) tells us that these theta-angles depend only
on $C_2$ and $C_4$. Intuitively this is what one would
expect. Indeed, after $T$ duality $C_4$ becomes $C'_0$ --- the new
RR zero-form, and  $C_2$ becomes $(C'_2)^{\vee}$. 
Our $Q_4$ becomes instanton number,
and $Q_0$ becomes the rank of the gauge group. The interaction with
the RR fluxes for the $1+5$ dimensional theory is of the form:
\begin{equation}\label{CommutativeTheta}
C_0\int_{{\bf R}^2}\trace {\cal F}+\int_{{\bf T}^4\times {\bf R}^{2}} 
\left(
C_2^{\vee}\wedge \mbox{tr} \; {\cal F}\wedge {\cal F}+
C_4 \; \mbox{tr} \; {\cal F}\wedge {\cal F} \wedge {\cal F}
\right)
\end{equation}
where ${\cal F}$ is the six-dimensional field strength.
Since these expressions are topological, they give theta-angles
after the dimensional reduction to $1+1$ dimensions\footnote{
The map from the worldsheet ${\bf R}^2$ to the moduli space of 
instantons on ${\bf T}^4$ implies the choice of the $U(N)$-bundle 
over ${\bf T}^4\times{\bf R}^2$. The theta-angles assign a phase 
to such a map, which is the linear combination of the Chern classes
specified in  (\ref{CommutativeTheta}).}.
Notice that the Ramond-Ramond zero-form $C_0$ couples only to
the $U(1)$ gauge field, which decouples 
from the other fields and does not participate in the
dual description of the AdS supergravity.

We explain the meaning of (\ref{CiH}) when $B\neq 0$ in the next section.

\section{Relation to the noncommutative torus.}
\subsection{Small torus and noncommutative torus (a very brief review).}
There is a beautiful relation between the noncommutative geometry and
the compactification of the string theory on the torus of very small size.
We will not review it here, since it is thoroughly explained 
in \cite{DH,CDS,SWNC}. But we will give a very brief 
description of this correspondence, in order to fix our notations.
Consider the compactification of the Type IIA string theory on the small torus 
${\bf T}^4$ with the metric $g_{ij}$ and the B-field $B_{ij}$. 
Consider  $N$ $D0$ branes in this background. 
Suppose that the size of
the torus is much less than the string length. Consider first the
case $B_{ij}=0$. Making T duality in all the four directions of
the torus, we get the theory of $N$ $D4$ branes on the dual torus
(which has very large size). According to \cite{Witten}, the low energy
worldsheet theory for these $D4$ branes is the $N=4$ supersymmetric
Yang-Mills theory. 
Now let us turn on some very small $B$ field. (In our notations,
$B$ field is dimensionless; for the string worldsheet wrapping
the cycle $(ij)$ of the torus, the $B$ field gives the phase factor
$e^{2\pi iB_{ij}}$ in the path integral). Since $B$ is small, we expect
that it does not considerably change the dynamics of $D_0$ branes. 
The effect of the small $B$ field is some small correction to
$N=4$ SYM on ${\bf T}^4$. It turns out, that this correction is precisely
turning on the noncommutativity parameter. Moreover, this result holds 
for arbitrary $B_{ij}$, not necessarily small. 

The precise correspondence
goes as follows. The noncommutative torus is defined in terms of the
algebra of functions on it 
(see \cite{KonechnySchwarz} and references therein for details). 
This is the noncommutative deformation
of the algebra of functions on the usual torus. If we denote
$\phi_i$ the coordinates on the torus, then the algebra of functions
is generated by $e^{i\phi_j}$ for $j=1,\ldots,4$. The algebra
of functions on the noncommutative torus may also be thought of 
as generated by $e^{i\phi_j}$, but instead of 
$e^{i\phi_j}e^{i\phi_k}=e^{i\phi_k}e^{i\phi_j}$ we have:
$$
e^{i\phi_j}e^{i\phi_k}=e^{2\pi i\Theta_{jk}}e^{i\phi_k}e^{i\phi_j}
$$
The bivector $\Theta_{ij}$ is called the noncommutativity parameter.
This defines the noncommutative torus ``as a manifold''.
To define the Yang-Mills functional, we have also to specify the metric.
It is done in the following way. One introduces a four-dimensional
 abelian algebra,
acting on the algebra of functions as derivations. This algebra
is the analogue of the algebra of constant vector fields on the
commutative torus. It remains abelian in the noncommutative case.
The metric on the noncommutative torus is defined as the metric $G^{ij}$
on this algebra (as on the linear space). We have
the following correspondence between the parameters of the
string theory torus ${\bf T}^4$ and the parameters of the noncommutative
torus $\dualtfour$:
\begin{equation}\label{ThetaG}
\Theta_{ij}=B_{ij},\;\;\; G^{ij}=(g^{-1})^{ij}
\end{equation}
The relation between these formulas and the effective open string 
metric/noncommutativity parameter of \cite{SWNC} is the following. 
The effective open string metric and
noncommutativity parameter are given by Eq. (2.5) of \cite{SWNC}:
\begin{equation}\label{Effective} 
G^{ij}=\left({1\over g+B}\right)_S^{ij},\;\;
\Theta^{ij}=\left({1\over g+B}\right)^{ij}_A
\end{equation}
Suppose that we are given the background with the metric $g_{ij}$ and
the B-field $B_{ij}$.  First, we do the T-duality
$g+B\to {1\over g+B}$. Then, we use (\ref{Effective}) to get 
$G={1\over g}$ and $\Theta=B$.

The vector bundle over the noncommutative torus is defined as a module
$\cal E$ over the algebra of functions. The gauge field is defined
as the set of covariant derivatives $\nabla_1,\ldots,\nabla_d$
acting in $\cal E$ in a way that agrees with the action of the algebra
of functions (see \cite{CDS} for details). The curvature 
$F_{ij}={[}\nabla_i,\nabla_j{]}$ is an element of $\mbox{End}\;{\cal E}$.
To understand the correspondence between the numbers of branes and the
topological numbers of the noncommutative Yang-Mills theory we have to 
review the noncommutative analogue of the Chern classes. 
In conventional (commutative) geometry, we can think of 
$\trace e^{F\over 2\pi i}$ as a differential form; its cohomology
class is the Chern character of our bundle. 
In the noncommutative case we cannot think of $\trace e^{F\over 2\pi i}$ 
as a differential form, in particular because the operation of integration
cannot be separated from the operation of taking the trace. 
We will think of it in the following way. Given the noncommutative
torus ${\bf T}^d$, consider a commutative torus $U(1)^d$, consisting of the
authomorphisms of the algebra $e^{i\phi_j}e^{i\phi_k}=e^{2\pi i\Theta_{jk}}
e^{i\phi_k}e^{i\phi_j}$ of the following form:
\begin{equation}
e^{i\phi_j}\to e^{i\alpha_j}e^{i\phi_j},\;\;\; \alpha_j\in[0,2\pi]
\end{equation}
Consider a trivial bundle over this torus with the fiber being
our module $\cal E$ (the infinite-dimensional space). 
Then, $F_{ij}={[}\nabla_i,\nabla_j{]}$ is 
a two-form on this commutative torus with values in 
$\mbox{End}\; \cal E$. Let us consider $\trace e^{F\over 2\pi i}$
as a (non-homogeneous) constant form on $U(1)^d$. This form by itself
does not have any integrality properties. However, it is known
from mathematical literature \cite{Eliott} that the following form
on $U(1)^d$:
\begin{equation}\label{Chern}
\mu=e^{\iota(\Theta)}\;\trace e^{F\over 2\pi i}
\end{equation}
is an integer form. It is the noncommutative
version of the Chern character for bundles on a torus.
The correspondence between the numbers of branes and the topological
numbers of the noncommutative Yang-Mills theory goes as follows.
For a two-form $\nu={1\over 2}\nu_{ij} dx^i\wedge dx^j$, let us denote
$\nu^{\vee}$ the two-form on the dual torus with the components 
$(\nu^{\vee})^{ij}={1\over 2}\epsilon^{ijkl}\nu_{kl}$.
Then, for $d=4$ 
we have the following relation between the Chern character and the
numbers of branes:
\begin{equation}\label{Mu}
\begin{array}{l}
\dss{\mu=\mu_0+\mu_2+\mu_4\;\mbox{vol},}
\vspace{5pt}\\
\dss{\mu_0=Q_0,\;\; \mu_4=-Q_4,\;\; \mu_2=Q_2^{\vee}}
\end{array}
\end{equation}
Indeed, to fix the relation between the 
components of $\mu$ and the charge $Z$, we can consider the case $B=0$.
After T duality, $C=C_0+C_2+C_4$ goes to 
\begin{equation}\label{Cprime}
C'=-C_4+C_2^{\vee}-C_0
\end{equation}
(it follows from (\ref{FifthVector}) and (\ref{CalC});
T duality acts on $\Gamma^{4,4}$ as $(x,v,y)\to (y,v^{\vee},x)$).
We use the coupling to the Ramond-Ramond fields to find $\mu$:
\begin{equation}
\int C'\wedge \mu =-Q_0C_0+(Q_2\cdot C_2)+Q_4C_4
\end{equation}
This implies (\ref{Mu}). 
In particular, we have the following relations:
\begin{equation}\label{TrOneAndTrF}
\begin{array}{c}
\dss{
\trace {\bf 1} = Q_0+(Q_2\cdot B)-{1\over 2}Q_4||B||^2}
\vspace{5pt}
\\
\dss{\trace {F\over 2\pi i}=\mu_2-\mu_4\Theta^{\vee}=Q_2^{\vee}
+Q_4 B^{\vee}}
\end{array}
\end{equation}
Also notice that for the geometrically dual torus, the K\"ahler forms
are related to the K\"ahler forms of the original torus as follows:
$$
\omega^i\mapsto(\omega^i)^{\vee}
$$
In our situation, we have Type IIB $Q_0$ $D1$, $Q_2$ $D3$ and $Q_4$ 
$D5$ branes.  In this case, we get the five-dimensional $U(Q_0)$ theory
on ${\bf R}^{1,1}\times \widetilde{\bf T}^4$, where ${\bf R}^{1,1}$
is commutative. (We may consider this as a particular case of 
the six-dimensional noncommutative Yang-Mills, with the noncommutativity
bivector $\Theta_{\mu\nu}$ vanishing when one of its indices is 
in the tangent space to ${\bf R}^{1,1}$.) 
Dimensional reduction of this theory on the four-torus gives 
the sigma-model with the target space the moduli space 
of instantons on  $\widetilde{\bf T}^4$. 
We will not discuss the details of the dimensional reduction
in this paper. Let us explain why the resulting metric on the
moduli space of noncommutative instantons is Hyper-K\"ahler.
The argument goes precisely as for the conventional (commutative)
instantons \cite{DonaldsonKronheimer}.
Consider the infinite-dimensional space of all connections
$\nabla_i$
in the module $\cal E$ over the algebra of functions on $\dualtfour$. 
This space is endowed with the flat metric 
$ds^2=\trace d\nabla_i d\nabla^i$
and the sphere worth of  K\"ahler forms 
$\Omega{[}\omega{]}=
\epsilon^{ijkl}\trace \omega_{ij}d\nabla_k \wedge d\nabla_l$,
$\omega$ being a Kahler form on $\dualtfour$.
There is an infinite-dimensional group of gauge transformations
$\nabla_i\to g\nabla_i g^{-1}$
preserving the metric and the complex structures.
Choosing a particular Kahler form $\Omega{[}\omega{]}$ 
as a symplectic form, we can think of the gauge transformation
with the infinitesimal parameter $\xi$ as generated by 
the Hamiltonian $H_{\xi}{[}\omega{]}=
\epsilon^{ijkl}\omega_{ij} \trace \xi F_{kl}$.
The anti-self-duality condition may be written as the condition
that for any Kahler form $\omega$
$$H_{\xi}{[}\omega{]}=(\omega\cdot\omega^+)\trace\xi$$ 
where
$\omega^+$ is the constant two-form on the right hand side of the
instanton equations $F^+=\omega^+$ \cite{ANS}. Therefore, the 
moduli space of the anti-self-dual connections is the Hyper-K\"ahler
reduction of the space of all connections. 
According to Theorem 3.2 in
\cite{HKLR}, the natural metric on the Hyper-K\"ahler quotient
is Hyper-K\"ahler.

As an example of how this correspondence works, let us compute the 
tension of the black string in the limit of the small size torus.
From the solution (\ref{UandTheta}), we see that the 
tension is:
\begin{equation}\label{tension}
T={1\over g_{str}}\sqrt{V^{\infty}||Z_+||^2}
\end{equation}
At small $V^{\infty}$, we have:
\begin{equation}
\begin{array}{rcl}
\dss{
||Z_+||^2}&\dss{=}&\dss{(Z\cdot \vec{v})^2+{(Z\cdot v^0)^2\over ||v^0||^2}
\simeq
||(Q_2-Q_4B)_+||^2+
{(||Z||^2-||Q_2-Q_4B||^2)^2\over 8 Q_4^2 V^{\infty}}+}\\
&\lefteqn{\dss{+{1\over 2}(||Z||^2-||Q_2-Q_4 B||^2)}}&
\end{array}
\end{equation}
(we have neglected the terms of the order $V^{\infty}$).
Then, the mass formula (\ref{tension}) gives:
\begin{equation}
\begin{array}{c}
\dss{T={1\over\sqrt{2}g_{YM}^2}\left\{
{\widetilde{V}\over 2Q_4}
\left|\; ||Z||^2-||Q_2-Q_4B||^2\right|
\pm\right.}\vspace{10pt}\\
\dss{\left.\pm
Q_4\left[1+
2{||(Q_2-Q_4B)_+||^2\over ||Z||^2-||Q_2-Q_4B||^2}\right]+
O(\widetilde{V}^{-1})\right\}}
\end{array}
\end{equation}
where $\pm$ is the sign of $||Z||^2-||Q_2-Q_4B||^2$ and 
$\widetilde{V}=(V^{\infty})^{-1}$.
The first term is leading 
when $V^{\infty}\to 0$. The second term after identifications
(\ref{ThetaG}) and (\ref{Mu}) is proportional to the
action of the noncommutative instanton, found in \cite{ANS}.

Now we are in a position to give a formula relating the near-horizon
geometry to the noncommutative torus. First, we look at which small size tori
at infinity flow to the given torus at the horizon. We use (\ref{AsvsNH})
with $V=0$. The third of relations (\ref{AsvsNH}) expresses $s$ in terms
of $V^h$. Substituting this $s$ to the second equation, we get the relation
between $B$ and $\omegr$. We have:
\begin{equation}
\begin{array}{l}
\dss{\lc=\lc^h,}\vspace{5pt}\\
\dss{B={Q_2\over Q_4}+{||Z||^2\over Q_4\sqrt{||Z||^2-2Q_4^2V^h}}\omegr^h-
{\sqrt{2V^h||Z||^2}\over \sqrt{||Z||^2-2Q_4^2 V^h}}\omegr}\vspace{5pt}
\end{array}
\end{equation}
These relations determine the family of small tori,
which flow to our attractor. This family is parametrized by 
$\omegr$. Each of these small tori determines the noncommutative torus,
which is geometrically dual to it, with the noncommutativity parameter
$\Theta=B$. Therefore, to the given attractor we associate the family
of noncommutative tori. This family is characterized by the fixed 
complex structure and the following relation between the K\"ahler form 
and the noncommutativity parameter:
\begin{equation}\label{Family}
\Theta={\mu_2^{\vee}\over \mu_4}+
{||\mu||^2 \over \mu_4\sqrt{||\mu||^2-2\mu_4^2V^h}}\omegr^h
-{\sqrt{2V^h||\mu||^2}\over \sqrt{||\mu||^2-2\mu_4^2V^h}}\Omegr^{\vee}
\end{equation}
(Here we have used $\mu$ instead of $Z$).
The formula (\ref{Family}) gives the answer to the question: which
noncommutative torus corresponds to the $AdS$ background 
with parameters $(\lc^h,\omegr^h, V^h, B^h)$? We see that the
answer is not unique. It follows that all the tori from the family 
(\ref{Family}) have the same moduli space of noncommutative instantons. 

Since the parameters of the attractor are unambiguously determined
by the parameters of the small torus $(B,\omegr,\lc)$, we can
rewrite the equations (\ref{Family}) for the family of noncommutative tori
so that it does not contain the parameters of the attractor. 
Indeed, rewriting (\ref{SmallVEquivalent}) in terms of $\mu$ and 
$(\Theta,\Omega^i)$, we get:
\begin{equation}\label{FamilyOfNCTori}
\begin{array}{cl}
\displaystyle{
1)} & 
\displaystyle{\Omegc=\mbox{const} \;\; \mbox{(modulo phase)}}
\vspace{5pt} \\
\displaystyle{2)} &
\displaystyle{ 
\Omegr-2{(\Omegr\cdot(\mu_2-\mu_4\Theta^{\vee}))\over 
||\mu_2-\mu_4 \Theta^{\vee}||^2-||\mu||^2}(\mu_2-\mu_4 \Theta^{\vee}) 
=\mbox{const}}
\vspace{5pt}
\end{array}
\end{equation}
Different constants give different families of noncommutative tori.
The form $\mu_2-\mu_4\Theta^{\vee}$ is expressed in terms of
$\trace F$ in (\ref{TrOneAndTrF}).
It would be interesting to prove in noncommutative geometry, 
that all the noncommutative tori from the family (\ref{FamilyOfNCTori})
have the same moduli space of instantons with instanton number $\mu$.

Notice that the Equations (\ref{FamilyOfNCTori}) imply 
that the self-dual part of the form 
\begin{equation}
{\mu_2-\mu_4\Theta^{\vee}\over  
||\mu_2-\mu_4 \Theta^{\vee}||^2-||\mu||^2}
\end{equation}
is a covariantly constant section of the restriction 
of the ``universal''  bundle
$\Lambda^2_+\widetilde{\bf T}^4$ of self-dual forms 
on the family of equivalent tori. 

Notice that for small noncommutativity and $\mu_2=0$, the second equation 
(\ref{FamilyOfNCTori}) implies that: 1) $\Omegr$ is constant
and 2) the self-dual part of $\Theta^{\vee}$ is constant.
This is what we intuitively expect. Indeed, in the limit
of almost commutative torus there are large regions in the moduli
space, where some instantons become very small. But very small instantons
on zero volume ${\bf T}^4$ is almost the same as instantons on ${\bf R}^4$.
According to \cite{SWNC}, the moduli space of instantons on ${\bf R}^4$
depends only on the self-dual part of the noncommutativity parameter. 
Some properties of supersymmetric configurations on ${\bf R}^4$ are
discussed in the next subsection.

\subsection{Supersymmetric configurations on ${\bf R}^4$.}
It was conjectured in \cite{NS,Berkooz} that the supersymmetric 
configurations on ${\bf R}^4$ with nonzero $B$ field are noncommutative
instantons. This has been proven in \cite{SWNC} in the
zero-slope limit, that is when $\alpha'\to 0$. We want to show that 
this is true beyond the zero slope limit.
The zero slope limit may be characterized by:
\begin{equation}\label{ZeroSlope}
(\alpha')^2 g^{ik} g^{jl} B_{ij} B_{kl} \simeq {1\over \epsilon}>>1
\end{equation}
Notice that
\begin{equation}
(\alpha')^2 g^{ik} g^{jl} B_{ij} B_{kl} =
{1\over (\alpha')^2} G_{ik} G_{jl} \Theta^{ij} \Theta^{kl}
\end{equation}
(this follows from the formula  (\ref{Effective}) for the effective 
open string metric and noncommutativity parameter).

We will argue that the moduli space of supersymmetric configurations
depends only on certain combination of metric and $B$ field, and
that for the background with generic $g$ and $B$ we can always 
find the background which gives the same moduli space of supersymmetric
configurations and satisfies the zero slope condition.
Let us prove it. Suppose that we are given 
the generic metric and $B$ field on ${\bf R}^4$, not necessarily
satisfying (\ref{ZeroSlope}). 
The moduli space of supersymmetric configurations 
should not change significantly if we replace our ${\bf R}^4$
by sufficiently large torus. (We will make this statement more precise 
below). This torus will have a very large volume
and a very large B-field; in fact, for the generic background the B-field
will scale like the square root of the volume.    The target space of
the sigma-model describing the low energy dynamics of the D1D5 system
wrapped on this torus is the moduli space of supersymmetric configurations
on this torus. For finite values of the charges, 
it has a region corresponding 
to supersymmetric configurations on ${\bf R}^4$. Indeed, for a very large 
torus we expect to have such field configurations that the field strength 
is localized in the small region of the torus which may be thought of
as ${\bf R}^4$ with a nontrivial $B$ field. We will call such field
configurations ``localized''. Since we have $g_{ij}\simeq B_{ij}$,
these field configurations do not satisfy the condition of the zero
slope limit. We want to argue, however, that our large torus is equivalent
to the torus with $|g_{ij}|<<|B_{ij}|$, for which the localized 
supersymmetric configurations do satisfy  the zero slope condition.   
Let us show it. It is very convenient to do first the T duality 
$g+B\to {1\over g+B}$
which gives us the torus with small $g_{ij}$ and $B_{ij}$.
The metric and the B field are, still, of the same order of magnitude.
In other words, we have: 
\begin{equation}\label{Regime}
||B||^2\approx V << 1,\;\; Q_0\simeq Q_4
\end{equation}
(and we put $Q_2=0$). Although the volume is small, it is finite. 
As we have 
argued in the previous section, we can always find an equivalent torus
with zero volume. The K\"ahler form $\omegr^0$ and the B-field
$B^0$ for that equivalent torus satisfy:
\begin{equation}\label{EquivInRegime}
B^0+s^0\omegr^0=B+s\omegr
\end{equation}
where
\begin{equation}
s^0=\sqrt{2V+s^2},\;\; s\simeq {Q_0\over Q_4}{1\over (B\cdot\omegr)}
\end{equation} 
(we have used an equation (\ref{s}) for $s$ in the regime (\ref{Regime})).
Since $s>>1$ and $V<<1$ we have:
\begin{equation}
s^0=s+O(V/s)
\end{equation}
Then, (\ref{EquivInRegime}) is satisfied by:
\begin{equation}
\omegr^0=\omegr,\;\; B^0=B+O(V/s)\simeq B
\end{equation}
(the correction for $B$ is much smaller then $B$ itself).
We see that the torus with small $V$ and $B$ satisfying (\ref{Regime})
is equivalent to the zero volume torus, which has the same shape and the
same (small) $B$ field. (Of course, we can take any other zero volume
torus with the same shape and the same self-dual part of the $B$ field).
In terms of the T-dual torus $(g_{ij}, B_{ij})$ with large $g_{ij}$ 
and $B_{ij}$ (the one we have started with) this means that it is equivalent
to the zero size torus with large B-field $\tilde{B}_{ij}$. However,
the relation between the shape of this equivalent torus and the shape of
$g_{ij}$, as well as the relation between $\tilde{B}$ and $B$ is more
complicated. In fact, it is very simple in terms of the effective
open string metric and the noncommutativity parameter. Namely, 
{\em the noncommutative torus
$(G_{ij}, \Theta^{ij})$ with small $\Theta^{ij}$ and 
$G_{ik}G_{jl}\Theta^{ij}\Theta^{kl}\simeq (\alpha')^2$ is equivalent
to the very large noncommutative torus ($G_{ij}=\infty$) 
with the same shape and the
same noncommutativity parameter}. The corresponding moduli space
is just the moduli space of instantons on large noncommutative ${\bf T}^4$
and the region in the moduli space corresponding to localized field
configurations is the moduli space of instantons on the noncommutative 
${\bf R}^4$. 

Let us summarize what we have. We have started with 
the torus of the large size and large B-field. We have argued that there
are localized field configurations, which are the same as supersymmetric
configurations on ${\bf R}^4$ with the $B$ field. On the other hand, we
have proven that the moduli space of supersymmetric configrations on this
torus is the same as the moduli space of supersymmetric configurations  
on the small size torus
with some other $B$ field (which is also large). This moduli space also has
a region corresponding to localized configurations. 
It is a natural conjecture
(we did not prove it) that the localized supersymmetric field
configurations on the large torus correspond to the localized configurations
on the equivalent small torus. But the localized configurations on the small
torus are described by the zero slope limit, and therefore they are 
instantons on the noncommutative ${\bf R}^4$.

This shows that the moduli space of supersymmetric configurations
on ${\bf R}^4$ does not depend on $\alpha'$. 

\subsection{Wilson Lines.}
The moduli space of anti-self-dual connections on 
${\bf T}^4$ has an obvious $U(1)^4$ symmetry:
\begin{equation}
\nabla_i\to\nabla_i+ia_i
\end{equation}
where $a_i$ are real numbers. This symmetry does
not have fixed points.  Therefore, the moduli
space has a structure of a bundle with the fiber four-torus 
$\bar{{\bf T}}^4$. In the commutative case (when $B=0$)
we could identify this ${\bf T}^4$ with  the dual to the
torus on which our Yang-Mills fields live. In the noncommutative
case we expect more subtle relation between these tori.
One way to see it is to notice that the periodic identifications
$a_i\equiv a_i+2\pi$ result from the gauge transformations,
which are modified for the noncommutative torus.
The gauge transformation with the parameter $e^{2\pi i \phi_i}$
would not just change $a_i\to a_i+2\pi i$, but will instead add
some nontrivial function of $\phi_1$ and $\phi_2$ to $\nabla_i$.
Shifting $a_i$ by constant will require more complicated gauge
transformation, and the periodicity properties of the Wilson
lines will be changed. We expect that the torus
$\bar{\bf T}^4$ parametrizing the Wilson lines will be 
different from the geometrically dual to ${\bf T}^4$.
Notice that in the presence of the $B$ field the shape of the torus
${\bf T}^4$ depends on the representative in the equivalence class of 
backgrounds. Therefore, if $\bar{\bf T}^4$ was just geometrically
dual to ${\bf T}^4$, we would get non-equivalent sigma-models for
equivalent backgrounds. In fact, we know from Section 4 that 
$\omega_{\bf C}$ and $B+s\omega_{\bf R}$ are invariants of the
equivalence relation. Therefore, it is natural to conjecture
that the Hyper-K\"ahler structure on the moduli space of flat connections
in the bundle with the given $(Q_4,Q_2,Q_0)$
is specified by the forms $\mbox{Re}\,\Omega_{\bf C}$,
$\mbox{Im}\,\Omega_{\bf C}$ and 
${\tilde{\Omega}_{\bf R}\over\sqrt{||\tilde{\Omega}_{\bf R}||^2}}$
where $\tilde{\Omega}_{\bf R}$ is given by the second line in 
(\ref{FamilyOfNCTori}).  We do not know how to prove
it in noncommutative geometry.

\subsection{Possible singularities in the moduli space.}
If the instanton moduli space for one of the tori
from the family (\ref{FamilyOfNCTori}) has singularities, then 
all the tori from this family have singular instanton moduli spaces. 
Singularities of the instanton moduli spaces are related 
in string theory to the possibility for the system of branes to
split into two or more subsystems preserving the BPS condition.
The conditions 
for the possibility of the decay of the black string have been worked
out in \cite{DoneDfive}. The answer is the following. For the black string
with the charge $Z$ in the background $W$ to be able to decay into
two strings, one with the charge $Z_1$ and the other with the 
charge $Z-Z_1$, it is necessary that the projection $(Z_1)_+$ of
$Z_1$ on $W$ and the projection $Z_+$ of $Z$ on $W$ are collinear vectors.
This means, that $W\cap Z_1^{\perp}=W\cap Z^{\perp}$. 
In other words, $W\cap Z^{\perp}$ should be orthogonal to $Z_1$.
Therefore, singularity or nonsingularity depends only on $Z^{\perp}\cap W$.
But our definition of equivalent tori is precisely that 
they have the same $Z^{\perp}\cap W$.  

We get the simplest example of the singular moduli space, if we
take $Q_2=0$, $B$ antiselfdual, and turn off the Ramond-Ramond fields. 
In this situation, $(B\cdot \omega)=0$ (this is the definition of
$B$ being antiselfdual), and the plane $W\cap Z^{\perp}$ is
generated by the three vectors $[(0,\omega^i,0),(0, 0)]$ together with
the fourth vector $[(0,0,0),(1,\kappa^{-2})]$. We see that the whole
lattice $\Gamma^{1,1}$ generated by the integer vectors of the form 
$[(*,0,*),(0,0)]$ is orthogonal to $W\cap Z^{\perp}$. Therefore,
our $(Q_4,Q_0)$ system may decay into $(Q_4',Q_0')$ and 
$(Q_4-Q_4', Q_0-Q_0')$, provided that $Q_4'Q_0'>0$ and
$(Q_4-Q_4')(Q_0-Q_0')>0$. The corresponding singularity in the
moduli space of instantons is due to small instantons. 
Notice that if we turn on the generic Ramond-Ramond fields,
then the separation of branes becomes impossible. Therefore,
although the moduli space of instantons remains geometrically singular,
the conformal sigma-model is nonsingular. The reason is, of course,
that we have turned on theta-angles. 

It is not true that an arbitrary background with singular instanton moduli
space is related to this example by dualities. Indeed, it can
happen that the sublattice orthogonal to $W\cap Z^{\perp}$ is not 
$\Gamma^{1,1}$ (and has dimension higher then two). Then, the corresponding 
singularity cannot be related to the small instantons by the chain of 
dualities.  

\subsection{Theta-angles.}
The target space ${\cal M}$ of the $D1D5$ sigma-model has a nontrivial 
second cohomology
group. There is a natural map $\theta_Z$ from the cohomology lattice 
of ${\bf T}^4$ to the second cohomology group of the moduli space
of noncommutative instantons with the charge $Z$. 
Let us remind how this map is 
constructed. Given a cohomology class $\nu\in H^2(\tfour, {\bf Z})$,
we want to describe how to compute $\theta_Z(\nu)$ on a cycle
$\Sigma\subset {\cal M}$. We describe the embedding 
$\Sigma\to {\cal M}$ as a connection  on a vector bundle 
$\widehat{\cal E}$ over $\Sigma\times {\bf T}^4$. Then,
the value of $\theta_Z(\nu)$ on $\Sigma$ is the highest component
of $\mu(\widehat{\cal E})\wedge \nu$. (This is the noncommutative
generalization of 
$\int_{\Sigma\times \tfour}ch(\widehat{\cal E})\wedge \nu$.)

Although $\theta_Z$ is naturally defined over ${\bf Z}$ we can
consider it over ${\bf R}$. Consider the fluxes of the Ramond-Ramond field
through the cycles of ${\bf T}^4$ as an element of 
$H^*({\bf T}^4, {\bf Z})$.
We want to show that the theta-terms in the $D1D5$ effective action
are the values of $\theta_Z$ on the {\em near-horizon} Ramond-Ramond
fields:
\begin{equation}
\mbox{Theta-angles}=\theta_Z(C^h)
\end{equation}
In particular, the theta terms do not depend on the choice
of the representative in the equivalence class of the backgrounds.

To derive this formula, we will use the following trick.
We first consider the case of  Euclidean
D-branes wrapped on ${\bf T}^6$. We will take this ${\bf T}^6$ to
be a product of ${\bf T}^2$ and a small ${\bf T}^4$, and go to
the dual noncommutative $\widetilde{\bf T}^4$.
In this case, we know that the Ramond-Ramond
fields couple to the winding numbers of various D-branes on this six-torus:
\begin{equation}\label{IntegerCoupling}
S_{RR}=C'\wedge\mu_{{\bf T}^2\times\dualtfour}
\end{equation}
where $\mu_{\ttwo\times\dualtfour}$ 
is the integer cohomology class of $\ttwo\times\dualtfour$
specifying the number of branes wrapped on various cycles. 
Then we look at (\ref{IntegerCoupling}) from the perspective of the
noncommutative geometry. We can explicitly express 
$\mu_{\ttwo\times\dualtfour}$
in terms of the field strength of the Yang-Mills field on 
$\ttwo\times\dualtfour$ using the Eliott's formula (\ref{Chern}):
\begin{equation}\label{RRforTsix}
\begin{array}{c}
\dss{
S_{RR}=C'\wedge\mu_{\ttwo\times\dualtfour}=
C'\wedge e^{\iota_{\Theta}}\left[\trace 
\exp\left({1\over 2\pi i}F\right)\right]
=}
\vspace{5pt}\\
\dss{=
(e^{\Theta}\wedge C)'\wedge\trace
\exp\left({1\over 2\pi i}F\right)=A'\wedge\trace
\exp\left({1\over 2\pi i}F\right)}
\end{array}
\end{equation}
(the forms $A$ and $C$ are defined in Appendix A, and 
$C'$ is given by (\ref{Cprime}).) 
Now we want to pass to ${\bf R}^2\times\dualtfour$ by making ${\bf T}^2$
very large, much larger then $\dualtfour$. The low energy theory is a 
dimensional reduction of the six-dimensional Yang-Mills on $\dualtfour$.
This theory lives on an ordinary, commutative two-torus. However,
it does remember, in a way, that it was obtained from the reduction
on the noncommutative manifold;
it turns out that the integral along this commutative two-torus of 
$\mbox{tr}\, F$ is not integer. This follows from the formula 
\begin{equation}\label{Eliott} 
\trace e^{{1\over 2\pi i}F}=e^{-\iota_{\Theta}}\mu
\end{equation}
derived in \cite{Eliott}.
For the worldsheet instanton $\mu$ has two indices along ${\bf T}^2$
and two or four indices along $\dualtfour$. 
Its contraction with the noncommutativity parameter gives nontrivial
(and noninteger) flux of $\mbox{tr}\, F$ along the two-torus. 
This flux contributes to the formula (\ref{RRforTsix}) for the coupling
to the RR fields. 

The appearence of noninteger $\int \mbox{tr}\, F_{z\bar{z}}$ 
may seem strange and we want to make a comment about it. 
Let us consider the following
example. Take our noncommutative torus ${\bf T}^4$ to be a product of two
noncommutative tori ${\bf T}_{(1)}^2\times {\bf T}_{(2)}^2$, with
the noncommutativity parameters $\theta_1$ and $\theta_2$. 
Let us consider an instanton with $\int_{{\bf T}^2\times {\bf T}^2_{(1)}}
F\wedge F \neq 0$ (the first ${\bf T}^2$ in the product 
${\bf T}^2\times {\bf T}^2_{(1)}$ is the commutative torus). 
In the adiabatic limit ({\it i.e.}, 
when the size of ${\bf T}^2$ is much larger then
the size of ${\bf T}^2_{(1)}$) this configuration may be thought of as 
wrapping the large (and commutative) ${\bf T}^2$ on the nontrivial 
two-cycle in the moduli space of flat connections in ${\bf T}^2_{(1)}$
(see \cite{BISV}). Naively it seems that we can just put $A_z$ and
$A_{\bar{z}}$ equal to zero, which would clearly give $F_{z\bar{z}}=0$,
and no flux. However, there is a subtlety here. In fact, we have to consider
flat connections modulo the gauge transformations. And it is not true that
for our map from ${\bf T}^2$ to the moduli space of flat connections
on ${\bf T}^2_{(1)}$ we can globally choose the representative in the 
gauge equivalence class smoothly depending on the point in $\ttwo$. 
Therefore, we should really cover our ${\bf T}^2$ with patches and construct 
separately the family  of
flat connections on ${\bf T}^2_{(1)}$ over each patch, and then glue them 
together by the
appropriate gauge transformations. These gauge transformations can make
it impossible to choose $A_z=0$ and $A_{\bar z}=0$. In the case if the small
torus is noncommutative, we have additional complications due to the
fact that our fields $A$ are not really gauge fields, but noncommutative
gauge fields. That is, they have non-standard gauge transformations. 
In fact, non-commutative gauge fields can be expressed in terms of the 
ordinary gauge fields, as explained in \cite{SWNC}. 
The formula to the first order in the noncommutativity parameter is:
\begin{equation}
\widehat{A_i}=A_i-{1\over 4}\Theta^{kl}(A_k(\partial_l A_i+F_{li})+
(\partial_l A_i+F_{li})A_k)
\end{equation}
The corresponding noncommutative field strength is, in our case:
\begin{equation}
\widehat{F_{z\bar{z}}}=F_{z\bar{z}}+
{1\over 4}\theta^{ij}\left[2(F_{zi}F_{\bar{z}j}+F_{\bar{z}j}F_{zi}-
(A_i(\nabla_j F_{z\bar{z}}+\partial_j F_{z\bar{z}})+
(\nabla_j F_{z\bar{z}}+\partial_j F_{z\bar{z}})A_i))\right]
\end{equation}
This expression shows that $\hat{F}_{z\bar{z}}$ cannot be taken to be zero. 
The noncommutative instanton is the deformation of the configuration with
$\int F\wedge F=8\pi^2 n_{inst}$ and $\int_{{\bf T}^2}F=0$. 
In this topological sector,
we can take $F_{z\bar{z}}=0$. However, $\hat{F}_{z\bar{z}}\neq 0$:
\begin{equation}
\int_{{\bf T}^4} \mbox{tr}\; \hat{F}_{z\bar{z}}=
{1\over 2}\int_{{\bf T}^4} \theta^{ij} F_{zi} F_{\bar{z}j}\simeq 
\theta_1 n_{inst}
\end{equation}
which is in agreement with (\ref{Eliott}).

Now let us see what happens when we replace ${\bf T}^2$ by ${\bf R}^2$,
which gives us our original configuration of $D1D5$ wrapped on
${\bf T}^4$  with two common noncompact directions. 
In this case, the flux of $\mbox{tr}\, F$ through ${\bf R}^2$ is not 
determined
by the topology. The configuration with zero $\mbox{tr}\; F_{z\bar{z}}$
minimizes the action. Indeed, we can add to the gauge field
$A_z$ the piece constant in ${\bf T}^4$: $A_z\to A_z+a(z,\bar{z})$,
$A_{\bar{z}}\to A_{\bar{z}}+\bar{a}(z,\bar{z})$, $a\in {\bf C}$.
The only modification to the action will be (remember that $\mbox{tr}$
includes integration along $\widetilde{{\bf T}}^4$):
$$\int_{{\bf T}^2}d^2z \;\mbox{tr}\; F_{z\bar{z}}^2 \to 
\int_{{\bf T}^2}d^2z \;\mbox{tr} \;(F-da)_{z\bar{z}}^2$$
Varying with respect to $a$, we get 
$(da)_{z\bar{z}}\;\mbox{tr}\;{\bf 1}=\mbox{tr}\;F_{z\bar{z}}$.
This means that the configurations with minimal action
have $\mbox{tr}F_{z\bar{z}}=0$. In particular, the flux of 
$\mbox{tr}\, F$
through ${\bf R}^2$ is zero. This implies, that the coupling to the
Ramond-Ramond fields becomes different from (\ref{RRforTsix}).
In fact, putting $\mbox{tr}F_{z\bar{z}}\to 0$ in (\ref{RRforTsix})
is equivalent to replacing the Ramond-Ramond fields $(C_0,C_2,C_4)$ 
by their near-horizon value $(C_0^h,C_2^h,C_4^h)$
given by the formula (\ref{CiH}). 
Let us prove it. We will consider the formula (\ref{RRforTsix}) valid for
the theory on the six-torus, but put the diagonal part of
the components of $F$ along ${\bf T}^2$ equal to zero. In other
words, replace $F$ with 
$F-\left({1\over \trace {\bf 1}}\int_{{\bf T}^2} \trace F\right)\sigma$
where $\sigma$ is the fundamental cohomology class of ${\bf T}^2$, 
$\int_{{\bf T}^2}\sigma=1$. We will get:
\begin{equation}\label{FtoZero}
\begin{array}{l}
\dss{\int_{{\bf T}^2}d^2z\left\{(e^{\Theta}\wedge C)'\wedge \trace 
\exp\left[{1\over 2\pi i}
\left( F-{\int_{{\bf T}^2}\trace{F}\over \trace {\bf 1}}\sigma
\right)\right]\right\}=}
\vspace{5pt}\\
\dss{=\exp\left[-{\int_{{\bf T}^2}\trace F\over 2\pi i \trace {\bf 1}}
\sigma\right]\wedge C'\wedge\int_{{\bf T}^2}d^2z\left\{
e^{\iota_{\Theta}}\;
\trace \exp\left[{F\over 2\pi i}\right]\right\}=}
\vspace{5pt}\\
\dss{=
\exp\left[-{\int_{{\bf T}^2}\trace F\over 
2\pi i\;\trace {\bf 1}}\sigma\right]\wedge
C'\wedge\mu_{\ttwo\times\dualtfour}=}
\vspace{5pt}\\ 
\dss{=C'\wedge\mu_{\ttwo\times\dualtfour}-
C'\wedge\mu_{\ttwo\times\dualtfour}
\wedge \left({\int_{{\bf T}^2} \trace F\over 2\pi i \;\trace {\bf 1}}
\sigma\right)=\hspace{10pt}}
\vspace{5pt}\\
\dss{=C'\wedge\mu_{\ttwo\times\dualtfour} - 
{(C'\cdot\mu_{\tfour})\over \trace\;{\bf 1}}\int_{\ttwo}\trace\; 
{F\over 2\pi i}
}
\end{array}
\end{equation}
Let us compare this expression with what we would get by replacing 
$C\to C^h$ in (\ref{IntegerCoupling}):
\begin{equation}\label{CtoCnh}
\begin{array}{c}
\dss{\int_{\ttwo}(e^{\Theta}\wedge(C-te^{-\Theta}))'
\wedge\trace \exp\left[{1\over 2\pi i}F\right]=}
\vspace{5pt}\\
\dss{=C'\wedge\mu_{\ttwo\times\dualtfour} 
-t \int_{\ttwo}\trace {F\over 2\pi i}}
\end{array}
\end{equation}
Substituting $t$ from (\ref{WhatIsTea}) and taking into account  
(\ref{TrOneAndTrF}) we see that (\ref{FtoZero})
is equal to (\ref{CtoCnh}). In other words, the coupling of
the sigma-model to the Ramond-Ramond fields is given by the
formula (\ref{IntegerCoupling}) with integer 
$\mu_{\ttwo\times\dualtfour}$ but $C$ replaced with $C^h$.
This means that the sigma-model on ${\bf R}^2$
couples to the Ramond-Ramond fields only through their near-horizon 
values.   

\section{Sigma-model Instantons and Supergravity Instantons.}

In this section we want to discuss the supergravity meaning of the
period map. The period map associates to each Hyper-K\"ahler manifold
the space of integrals (periods) over its two-cycles of
the three K\"ahler forms. Given the period map of the target space, 
one can immediately
compute the action of any sigma-model instanton. Indeed, the corresponding
map $f:{\bf R}^2\to {\cal M}$ from the wordsheet to the target space 
should be holomorphic in some complex structure and the action is given by 
\begin{equation}\label{WSInstAction}
S=\sqrt{\sum_{I=1}^3\left(\int_{{\bf R}^2} f^*\omega_I\right)^2}
\end{equation}
But the integrals over the worldsheet of the pullbacks of the 
K\"ahler forms is given in terms of the period map and the topology
of the instanton. More exactly, let us introduce the cohomology lattice
$H^2({\cal M},{\bf Z})$. There is a natural nondegenerate pairing on
this lattice (the construction is reviewed in \cite{Dijkgraaf}). 
The topology of the sigma-model instanton can be specified by giving
a vector $z\in H^2({\cal M}, {\bf Q})$, 
so that the integral of $p\in H^2({\cal M}, {\bf Z})$
over the worldsheet equals $(z\cdot p)$. 
The period map gives us a three-plane 
$W_3^{\sigma-model}\in {\bf R}\otimes H^2({\cal M}, {\bf Z})$,
and the action of the instanton is
\begin{equation}\label{Action}
S=\sqrt{||P_{W^{\sigma-model}_3}\;z||^2}
\end{equation}
We want to interpret this formula in supergravity.
Our sigma-model arises on the worldvolume of the $D1D5$ system
wrapped on ${\bf T}^4$. Let us consider the $D1D5$ system with
the worldvolume not ${\bf R}^2$, but some compact Riemann surface $\Sigma$
(for example, $S^2$). Of course, this system
will not be BPS and therefore will presumably collapse and decay into
gravitons and other fields of six-dimensional supergravity.
But let us consider the case when we have nontrivial instanton charge
on $\Sigma$. This instanton charge will couple to the Ramond-Ramond background
fields, and therefore it cannot decay into the states from the perturbative
string spectrum. In fact it decays into some perturbative state plus one
of the D-instantons of the six-dimensional sueprgravity, such as
Type IIB D-instanton, or
Type IIB Euclidean D-string wrapped on the two-cycle in ${\bf T}^4$,
or Type IIB Euclidean D3 brane wrapped on ${\bf T}^4$, or a combination
of them. Therefore, we can think of D-instantons as black strings with
compact worldsheet wrapping a nontrivial cycle in the target space.

Do we expect the action of these supergravity instantons to be equal to
the action of the corresponding instantons in our sigma-model? 
For generic backgrounds, we do not. One reason is that we expect the sigma-model
description to break down for small enough worldsheets. But even if we forget
about the possible breakdown of the low energy description, there is still a reason 
why the action of supergravity instantons is different. As we have 
explained in section 5.3, 
the sigma-model instanton on ${\bf R}^2$ usually has nonzero 
abelian gauge field strength on ${\bf R}^2$, determined from minimizing
the action. 
In generic situation, the flux of this gauge field through 
${\bf R}^2$ is non-integer.  If we want to compactify ${\bf R}^2$ to $S^2$
or any other Riemann surface,
we have to change the two-dimensional abelian gauge field so that the
flux is integer. This will increase the action. Therefore, we expect
that in the generic background, the actions of the supergravity instantons
are different from the actions of the sigma-model instantons. 
However, for some special backgrounds and instanton charges, the
flux of the abelian field happens to be zero. The condition for
it is that the coupling of the instanton to $(C_0, C_2, C_4)$ is
the same as the coupling to $(C_0^h, C_2^h, C_4^h)$.  In other
words, the charge $\chi$ of the supergravity instanton, defined so that
the phase factor is $e^{2\pi i ({\cal C}\cdot \chi)}$, satisfies:
\begin{equation}\label{MutualBPS}
(P_{W_4}Z\cdot \chi)=0
\end{equation}
We will see that this condition has a natural interpretation in supergravity.
We will argue that if this condition is satisfied (together with some other conditions) 
then  the action of the supergravity
instanton is equal to the action of the sigma-model instanton.

\subsection{Can supergravity instantons be ``absorbed'' by branes?}
We will argue that under certain conditions (including (\ref{MutualBPS}))
there exists the solution of the
Euclidean six-dimensional supergravity, corresponding to the black
string with the worldsheet ${\bf R}^2$ and the D-instanton at some 
distance from it. This solution preserves four real supercharges.
It has moduli corresponding to the motion of the D-instanton
in the space transverse to the black string. The action does not depend
on the position of the D-instanton. It is possible
that such a D-instanton can approach our black string and ``dissolve''
into  it, becoming a sigma-model instanton. We conjecture that the action
is not changed in this process. For example, we may consider the black
string which is obtained by wrapping $N>1$  Type IIB $D3$ branes on
some two-cycle of the torus. If we turn off the B-field on the torus,
then introducing a $D_{-1}$-brane will leave eight supersymmetries
unbroken. The $D_{-1}$-brane can move in the directions transversal
to $N$ $D3$, or it can become an instanton on the 
worldvolume of $N$ $D3$. 
From the point of view of the Yang-Mills theory on the worldvolume
of the threebranes, we can describe this process as follows. 
The $D_{-1}$ brane
on top of $N$ $D3$ branes may be thought of as point-like instanton.
This point-like instanton corresponds to the special point in the
moduli space of instantons of $SU(N)$ theory. Moving in the moduli
space, we deform it to the finite size instanton, which 
can be described as a classical supersymmetric field configuration on the
worldsheet. 

We should stress, however, that it is not always true that
the supersymmetric configuration of the supergravity instanton and the
D-brane can be represented as a classical instanton on the worldvolume
of the D-brane. For example, the configuration consisting of a single 
$D3$ brane and a D-instanton on top of it is supersymmetric; 
nevertheless, there are no
finite size instantons in the $U(1)$ gauge theory living on the worldvolume
of the single $D3$ brane. 
For our configuration consisting of the black string and the instanton, 
we do not know
whether the instanton can really be represented by the smooth 
worldsheet field 
configuration\footnote{Supersymmetric solutions of Yang-Mills equations
 on six-dimensional manifolds have been studied in
\cite{CDFN,Ward,AFFS,BlauThompson} and references therein.
For these solutions, the field strength should
belong to $su(3)\subset su(4)=so(6)$ modulo constant. 
(The variation of fermion with nonzero constant field strength can be
compensated by the $\eta^*$ transformation, as explained in 
\cite{SWNC}.)}. We can probably say that there is always an ``ideal''
instanton represented by the singular fields, but the precise 
meaning of this statement is not very clear. 

To prove the connection between the supergravity instantons
and the sigma-model instantons, we should answer two questions.
The first question is whether the supergravity instanton sitting on top of 
the brane can be deformed to the smooth supersymmetric solution of the
six-dimensional Yang-Mills equations.
In other words, whether the corresponding singular field configuration is 
just a point on the boundary space of smooth configurations. 
The second question is whether this smooth six-dimensional solution 
(if it
exists) has a good limit when we shrink ${\bf T}^4$ to zero size; if yes,
then it should correspond to the instanton of the $D1D5$ sigma-model.

We did not prove that the answers to these two questions are positive.
We want to give an example which shows that at least the first 
assumption is true for some instantons.
Some supersymmetric solutions of the six-dimensional Yang-Mills theory on 
Calabi-Yau
threefold were obtained (in the language of complex geometry) in
\cite{FMW}. Our example is analogous to the solutions found in that paper.

Consider Type IIB compactifyed on a six-torus  $\tsix=\ttwo\times\tfour$.
Introduce the following brane configuration. 
Take $N$ $D3$ branes wrapped on $\tfour$ and one 
$D3$ brane wrapped on $\ttwo\times\alpha$ where 
$\alpha\in H_2(\tfour,{\bf Z})$, $||\alpha||^2>0$.
The four-torus $\tfour$ with a given metric admits a family 
of complex structures parametrized by $S^2$. We can choose 
one of these complex structures so that the cohomology class
Poincare-dual to $\alpha$ (which we will call $[\alpha]$)
is of the type $(1,1)$. Then, there are holomorphic line bundles
on $\tfour$ whose Chern class is $[\alpha]$. Let us choose one of them
and call it ${\cal L}$. The other bundles with the same Chern class
may be obtained from this one by shifts in $\tfour$.
It is proven in Chapter 2.6 of \cite{Griffiths} that 
$$\mbox{dim}\, H^0(\tfour,{\cal L})={||\alpha||^2\over 2}$$
Zeroes of the global sections of ${\cal L}$ are holomorphic 
curves representing $\alpha$. 
Fixing one particular bundle corresponds to fixing two out
of ${||\alpha||^2\over 2}+1$ complex moduli of such holomorphic
curves. Let us parametrize the global sections of ${\cal L}$ by
the vectors $v\in {\bf C}^{||\alpha||^2\over 2}=H^0(M,{\cal L})$.
We will denote the corresponding sections $\phi[v](x_1,x_2)$, where
$x_1$ and $x_2$ are the local complex coordinates on $\tfour$.
Our configuration of $D3$ branes can be represented by the following
equation in $\tsix=\ttwo\times\tfour$:
\begin{equation}\label{Singular}
F_0(x_1,x_2,x,y)\stackrel{def}{=}
\phi[v_0](x_1,x_2)(a_0+a_1x+a_2y+\ldots + a_N x^{n/2})=0
\end{equation}
Here $(x,y)$ is the point of $\ttwo$ described by the Weierstrass
equation  $y^2=x^3+ax+b$.
If $N$ is odd, then the last term is $x^{N-3\over 2}y$. 
This equation represents a union of $N$ $D3$ branes 
wrapping $\tfour$ and the single $D3$ brane wrapping 
$\ttwo\times\alpha$. It preserves $1\over 8$ of supersymmetries.
One can think of this equation as specifying a single complex
surface in $\tsix$, but then this surface is singular.
To make it nonsingular, we consider a deformation of (\ref{Singular})
specifyied by the choice of
\begin{equation}
[u_0,\ldots, u_N]\in 
{\bf P}({\bf C}^{N+1}\otimes H^0(\tfour,{\cal L}))
\end{equation}
This deformation is:
\begin{equation}\label{Deformed}
\begin{array}{l}
F[u_0,\ldots,u_N](x_1,x_2,x,y)=\\ \vspace{5pt}=
\phi[u_0](x_1,x_2)+\phi[u_1](x_1,x_2)x+\phi[u_2](x_1,x_2)y+\ldots=0
\end{array}
\end{equation}
In particular, $F[a_0v_0,\ldots, a_Nv_N]=F_0$.
Unlike the initial equation (\ref{Singular}), it represents 
a smooth four-surface $X[u_0,\ldots,u_N]$
in $\tsix$. This smooth four-surface belongs to the same homology class 
$N[\tfour] + \alpha\times[\ttwo] $ as the 
original singular surface (\ref{Singular}).
Let us now assume that $N$ is even and $\alpha$ is divisible by two.
Then, the Chern class of the normal bundle to $X$ is divisible by two,
and $X$ is a spin manifold.
Therefore, we can wrap a Euclidean
three-brane on it \cite{KTheory,FreedWitten}. 
Let us take $n>1$ such $D3$ branes on top of each other and put $k$ 
$SU(n)$ instantons on their world-volume (the corresponding
smooth field configuration exists for large enough $k$ \cite{Taubes}).

After making T-duality in both directions of $\ttwo$, we get a system of 
$nN$ $D5$ branes with nontrivial field configuration on it. 
Since the initial field configuration was nonsingular,
we should get nonsingular configuration after T-duality. 
The brane charge of this configuration is 
$$ Nn [\tsix]+
\left(k-{N||\alpha||^2\over 8}\right) [\tfour] + 
n [\ttwo]\wedge \alpha$$
The term $N||\alpha||^2\over 8$ appears because $X$ has notrivial topology,
and therefore our $D3$ brane couples to axion even if we turn off the
world-volume gauge fields. Namely, the  D-instanton charge induced by 
the nontrivial topology of $X$ is 
$$
-\hat{A}(TX)[X]={p_1(TX)\over 24}[X]=-{1\over 24}\int_{\tsix} 
[X]\wedge [X]\wedge [X]=-{1\over 8}N||\alpha||^2
$$ 

This example demonstrates that at least some configurations exist as smooth 
solutions on the brane worldvolume.
However, the configuration which we considered is obviously
not the most generic one.
The most serious restriction in this example is that 
$$
\int_{\tsix} \trace F\wedge F\wedge F=0
$$
(this is what allowed us to reduce dimension from six to four by doing
T duality). Also, we did not prove that turning on the $B$ field will not 
destroy our configuration. 
   
In the remaining part of this section we will just assume that supergravity 
instantons
correspond to smooth field configurations and are well described by the
$D1D5$ sigma-model, and see what this assumption impies.  

\subsection{Classification of instantons in six-dimensional supergravity.}
The simplest instanton to consider is the $D_{-1}$-instanton, which is the 
dimensional reduction of the D-instanton in ten-dimensional Type IIB. 
Its action is
\begin{equation}\label{DMinusOne}
S_{D_{-1}}=C_0 - {i\over  g_{str}}
\end{equation}
The real part of this action is the phase associated to the instanton.
It is equal to the expectation value of the axion field. In the classical
low energy supergravity we have a symmetry $C_0\to C_0+\mbox{const}$. 
This  symmetry is a subgroup ${\bf R}^1$ of $SO(5,5,{\bf R})$. 
But the higher derivative corrections to the low energy action do not
have such a symmetry. The reason is precisely the contribution of 
the $D_{-1}$-instantons, which carry a phase depending on $C_0$.
Now we want to consider more general instantons. Notice that 
the subgroup ${\bf R}^1\subset SO(5,5,{\bf R})$ consisting of the shifts
of axion can be represented in terms of the light-like vector
$v\in {\bf R}\otimes \Gamma^{5,5}$, and the orthogonal vector 
$\chi\in v^{\perp}/v$. Indeed, let us consider the following $v$ and 
$\chi$:
\begin{equation}
\begin{array}{c}
v=[(0,0,0),(0,1)] \\
\chi=[(1,0,0),(0,0)]
\end{array}
\end{equation}
One can check that the one-parametric group of transformations
$T_{v,\chi}(t)\in SO(5,5,{\bf R})$ acting on vectors 
$x\in {\bf R}\otimes\Gamma^{5,5}$ in the following way:
\begin{equation}\label{Shifts}
T_{v,\chi} (t) x = x + t\left[(\chi\cdot x) v - (v\cdot x) 
\left(\chi+{t\over 2} ||\chi||^2 v\right)\right]
\end{equation}
is precisely the group of shifts of $C_0$.
We can consider arbitrary pairs $(v,\chi)\in \Gamma^{5,5}$, such that
$v$ is lightlike and $\chi$ is orthogonal to it. Any pair $(v,\chi)$
specifies the ${\bf R}$-subgroup of $SO(5,5,{\bf R})$, and there is 
a corresponding instanton breaking this symmetry. To find the action of
this instanton, we have to rewrite (\ref{DMinusOne}) in terms
of $v$ and $\chi$:
\begin{equation}\label{Covariant}
S={(\chi\cdot P_Wv)\over ||P_Wv||^2}-
i \sqrt{||P_{W\cap v^{\perp}}\chi||^2\over ||P_W v||^2}
\end{equation}
Invariance under U-duality group $SO(5,5,{\bf Z})$ implies that 
(\ref{Covariant}) gives the action of the six-dimensional instanton
for an arbitrary pair $(v,\chi)$. 
Although the $D_{-1}$-instanton we have started with was somewhat 
special because
the corresponding vector $\chi$ was a null-vector, 
the expression (\ref{Covariant}) is valid for all D-instantons.
Indeed, let us consider the configuration consisting of $Q_{3}$ Euclidean
$D3$-branes wrapped on ${\bf T}^4$, and $Q_{-1}$ $D_{-1}$-instantons. 
If the $B$-field
on the torus is anti-self-dual, this action for this configuration
is the sum of the action of $Q_{-1}$ $D_{-1}$-instantons and $Q_3$ 
Euclidean
$D3$-branes. This is in agreement with (\ref{Covariant}). The backgrounds
with a generic $B$-field may be obtained from this one by the rotation
by the element of $SO(4,4,{\bf R})$ preserving the vector 
$\chi=(Q_3,0,Q_{-1})$. The formula (\ref{Covariant}) is invariant
under such rotations. This means that (\ref{Covariant}) is valid for
an arbitrary $\chi$ of the form $(Q_3,0,Q_{-1})$. But any other primitive 
vector in the lattice $\Gamma^{4,4}$ is equivalent to one of these.

We will actually consider only D-instantons, therefore all that 
we need is (\ref{Action})  for $v=[(0,0,0),(0,1)]$.
In this case, $||P_Wv||^2=\kappa^2$, and (\ref{Covariant}) becomes:
\begin{equation}\label{ActionForD}
S=({\cal C}\cdot\chi) - {i\over \kappa}\sqrt{||P_{W_4}\chi||^2}
\end{equation}
(We denote $W_4=W\cap v^{\perp}$). 

\subsection{When do supergravity instanton and black string preserve
supersymmetry?}
So far we have discussed instantons
in flat six-dimensional space. Now let us introduce the black string with
some charge vector $Z$. In the presence of this black string, the
group $SO(5,5,{\bf R})$ of symmetries of the classical supergravity
is broken to the subgroup $SO(4,5,{\bf R})$ preserving the vector $Z$.
In particular, the subgroup of shifts $T_{v,\chi}$ defined in (\ref{Shifts})
preserves $Z$ if and only if $(Z\cdot \chi)=0$ (we are
considering only ``D black strings'', that is $(Z\cdot v)=0$). 
Therefore, if $(Z\cdot \chi)=0$, it makes sense to consider the
instantons ``charged'' under this symmetry. The condition $(Z\cdot \chi)=0$
by itself does not guarantee that the instanton in the presence of the
black string is a supersymmetric configuration. For example, let us consider
the $D_{-1}$-instanton in the background of the $D3$-brane wrapped
on the two-cycle of the torus. If there is a nonzero flux of the
$B$ field through this two-cycle, then this configuration breaks
all the supersymmetry.  The condition for this configuration to be
supersymmetric is that the flux of $B$ through the two-cycle on which
we wrap our $D3$ brane is zero. This means that 
\begin{equation}\label{MBPS}
(P_{W_4}\chi\cdot Z)=0
\end{equation}
In general case,
the condition (\ref{MBPS}) is a necessary condition for the instanton
and the black string to be a supersymmetric configuration. 
One can see it in the following way. Let us consider the instanton
in the presence of the large number $N$ of coinciding black strings, 
each having charge $Z$.
At each point of the transverse space, the instanton feels 
approximately flat background. Therefore, its action is given approximately
by (\ref{ActionForD}) but with $W$ depending on the distance from the black
string, according to the attractor equation.
The conditions for the action (\ref{ActionForD}) to be constant when
$W$ changes according to the attractor equation with given $Z$
are precisely that $(P_{W_4}Z\cdot \chi)=0$ and $(Z\cdot \chi)=0$.
We will prove in Appendix B that the configuration consisting of
the black string and the instanton preserves supersymmetry if
the following conditions are satisfied: 
\begin{equation}
\begin{array}{lc}
1) & (Z\cdot \chi)=0\\
2) & (Z\cdot P_{W_4}\chi)=0\\
3) & ||\chi||^2\geq 0
\end{array}
\end{equation}

\subsection{Supergravity instantons and period map.}
When the condition (\ref{MBPS}) is satisfied, we can rewrite 
$||P_{W_4}\chi||^2$ as follows:
\begin{equation}\label{IfDisBPS}
||P_{W_4}\chi||^2=||P_{W_4\cap Z^{\perp}}\chi||^2
\end{equation}
It makes sense to think of D-instantons classically when their action 
is very large, that is $g_{str}<<1$. 
However, nothing prevents us from taking simultaneously  
$g_{str}\sqrt{||Z||^2}>>1$. Therefore, there is a regime where we can
trust both the formula for the action of the D-instanton and the
near-horizon AdS supergravity. 

How can D-instantons help us to prove that the geometry of the 
target space does not receive string loop corrections? 
We know that the target space is a Hyper-K\"ahler manifold
with the second cohomology group isomorphic to 
$\Gamma^{4,4} \cap Z^{\perp}$.
The period map can be described as a choice of the 
positive-definite three-plane 
$W_3^{\sigma-model}\subset Z^{\perp}\cap ({\bf R}\otimes \Gamma^{4,4})$.
It was conjectured in \cite{Dijkgraaf} that 
\begin{equation}\label{WThree}
W_3^{\sigma-model}=Z^{\perp}\cap W_4
\end{equation}
We want to check this formula by comparing the actions of
the sigma-model instantons to the actions of the supergravity instantons.
We have argued that for some special backgrounds the sigma-model
instantons are related to the D-instantons in supergravity.
The actions of these supergravity instantons are given by 
(\ref{ActionForD}). We expect the action of the supergravity
instanton with the charge $\chi\in Z^{\perp}$ to be equal to the action 
of the corresponding  sigma-model instanton when 
$(\chi\cdot P_{W_4}Z)=0$. In other words, when $\chi$ is orthogonal
to the projection of $Z_+=P_{W_4}Z$ on $Z^{\perp}$. Let 
$L\subset {\bf R}\otimes \Gamma^{4,4}$ be 
the six-dimensional space of all vectors in $Z^{\perp}$ which are 
orthogonal to $P_{W_4}Z$:
\begin{equation}
L=({\bf R}\otimes \Gamma^{4,4})\cap Z^{\perp}\cap (P_{W_4}Z)^{\perp}
\end{equation}
Then, our condition on the instanton charge is just: $\chi\in L$.
For generic background there are no such integer vectors $\chi$
which satisfy this condition. However, there is a dense set of such 
backgrounds that we can find $\chi\in L$. Moreover, for any vector 
$v\in L$
we can find an integer vector $V\in\Gamma^{4,4}$ and a number 
$\lambda\in{\bf R}$
such that $\lambda V \approx v$. This means (assuming the smooth 
dependence of $W_3^{\sigma-model}$ on the background fields) 
that the action formula
for the supergravity instantons gives us the angle between
the plane $W_3^{\sigma-model}$ and an arbitrary vector in $L$.
Namely, given $v\in L$, the action formula (\ref{ActionForD})
implies that:
\begin{equation}
||P_{W_3^{\sigma-model}}\; v||^2=||P_{W_4}\; v||^2
\end{equation}
For $v\in Z^{\perp}\cap W_4$ we have 
$||P_{W_3^{\sigma-model}}\; v||^2=||v||^2$. This implies (\ref{WThree}).
Therefore, the action formula for supergravity instantons is in agreement
with the conjectured period map. In particular, we see that the periods do 
not depend on the string coupling constant. 

We should notice, however, that the validity of this argument depends 
on the assumption made at the end of Section 6.1. 

\bigskip

{\bf Acknowledgements.} I would like to thank Prof. E.~Witten for
many interesting and helpful discussions. Conversations with 
M.~Krogh and N.~Nekrasov were very useful. This work has been
supported in part by the Russian grant for support of scientific schools
No. 96-15-96455.

\appendix
\section{Moduli of $K3$ or ${\bf T}^4$ as points of Grassmanian.}
Here we will review the correspondence between the 
moduli and the points of the Grassmanian manifold, following
mostly \cite{Aspinwall}.
We will concentrate on the case of $K3$, the case of ${\bf T}^4$ is
similar. 

Let us first turn off the RR fields and consider weak coupling limit.
There is a one to one correspondence between the
points of the moduli space and the positive 4-planes
in ${\bf R}^{4,20}$. The moduli include the metric and the $B$ field.
First, let us describe the shape of $K3$. Let $\Pi\subset {\bf R}^{4,20}$
be a positive 4-plane. Choose $w\in\Gamma^{4,20}$ --- a primitive light-like
vector. Notice that
\begin{equation}
(w^{\perp}\cap \Gamma^{4,20})/w \simeq \Gamma^{3,19}
\end{equation}
We can consider the three-plane $\tilde{\Sigma}=\Pi\cap w^{\perp}$ 
and $\Sigma=p(\tilde{\Sigma})$ where $p:w^{\perp}\to w^{\perp}/w$
is the projection. The 3-plane $\Sigma$ corresponds, via the period
map, to the Hyper-K\"ahler structure of our $K3$. 
Now we want to specify the volume and the $B$ field. To do this, let us choose
another primitive light-like vector, $w^*$, so that $(w\cdot w^*)=1$.
We represent $\Pi$ as an orthogonal direct sum:
\begin{equation}
\Pi=\tilde{\Sigma}\oplus {\bf R}\tilde{B},\;\;\; \tilde{B}\perp\tilde{\Sigma}
\end{equation}
with
\begin{equation}
\tilde{B}=\alpha w +w^*+B
\end{equation}
and
\begin{equation}
(B\cdot w)=(B\cdot w^*)=0
\end{equation}
We identify $B$ as the $B$ field, and $\alpha$ as 
$V-{1\over 2}(B\cdot B)$. 
We have an equivalence $B\approx B+\nu$, with $\nu\in H^2(K3,{\bf Z})$.
Such a shift of $B$ can be accounted for as an ambiguity
in the choice of $w^*$. Indeed, let us identify the cohomology classes
of $K3$ with the vectors in the lattice $\Gamma^{3,19}$ which is the 
sublattice in $\Gamma^{4,20}$, orthogonal to both $w$ and $w^*$.
Given such a vector $\nu$, let us consider the following automorphism of
the lattice:
\begin{equation}
\begin{array}{c}
w^*\mapsto (w^*)'=w^*-\nu-{1\over 2}w(\nu\cdot\nu),\\
\mu\mapsto \mu'=\mu+(\mu\cdot\nu)w,\\
w\to w
\end{array}
\end{equation}
where $\mu$ is any vector in $\Gamma^{3,19}$. Then the new $w$ and $w^*$ are
orthogonal to the new $\Gamma^{3,19}$. If we also change $B$ and $\alpha$ as
follows:
\begin{equation}\label{transformation}
\begin{array}{c}
B\mapsto \left(B+w(B\cdot\nu)\right)+
\left(\nu+w(\nu\cdot \nu)\right)\\
\alpha\mapsto \alpha+{1\over 2}\left[||B||^2-||B+\nu||^2\right]
\end{array}
\end{equation}
then the new $\tilde{B}$ coincides with the old $\tilde{B}$, 
therefore all that we have 
done was to pick different $w^*$. The transformation laws 
(\ref{transformation}) suggest that $B$ should be identified with 
the $B$ field, and $\alpha$ should be identified as follows:
\begin{equation}
\alpha=V-{1\over 2}(B\cdot B)
\end{equation}
As far as we know, this formula for $\alpha$ was first obtained
in \cite{DanAndSanjaye}.
Following \cite{Dijkgraaf}, we use the notation  $(x, v, y)$ for the vector
$xw^*+yw+v$, where $x,y\in {\bf R}$ and $v\in {\bf R}^{3,19}$.

Now let us include the Ramond-Ramond fields and the coupling constant. 
We enlarge our lattice from $\Gamma^{4,20}$ to $\Gamma^{5,21}$ by
adding $\Gamma^{1,1}$. We will denote the vectors from ${\bf R}^{5,21}$
in the following way: $[(x,v,y),(x',y')]$ where $(x,v,y)$ is
the vector from ${\bf R}^{4,20}$ and $(x',y')$ is the vector from 
${\bf R}^{1,1}$. The scalar product is:
$$
||\;[(x,v,y),(x',y')]\;||^2=2xy+||v||^2+2x'y'
$$
Now we are ready to describe the correspondence between the moduli 
and the points of Grassmanian. The plane $W$ is generated by the following
five vectors.
Four vectors are of the form 
\begin{equation}\label{FourVectors}
E^{\mu}=[v^{\mu}, (0,-({\cal C}\cdot v^{\mu}))]
\end{equation}
with $v^i=(0,\omega^i,-(B\cdot\omega^i))$ and 
$v^0=(1,B, V-{1\over 2}(B\cdot B))$, and the fifth vector is
\begin{equation}\label{FifthVector}
E^5=\left[ {\cal C}, 
\left(1\;, \; {1\over \kappa^2}-{||{\cal C}||^2\over 2}\right) \right]
\end{equation}
Let us explain what is $\kappa$ and ${\cal C}$.
We define ${\cal C}\in {\bf R}^{4,20}$ as follows:
\begin{equation}\label{CalC}
{\cal C}=(-C_0, C_2, C_4)
\end{equation}
where $C_0,C_2,C_4$ are related to the Ramond-Ramond fluxes $A$ 
in the following way:
\begin{equation}\label{CvsRR}
C=C_0+C_2+C_4=(A_0+A_2+A_4) e^{-B}
\end{equation}
Here we consider $A_0,\; C_0$ zero-forms, $A_2,\; C_2$ and $B$ 
two-forms and $A_4,\; C_4$ four-forms. The coupling of the fields
on the brane world-volume to the Ramond-Ramond fluxes is:
\begin{equation}
\int A\wedge \trace \exp\left({1\over 2\pi i}F-B\right)
\end{equation}
The formula (\ref{CvsRR}) can be verified by considering the 
transformations  of ${\cal C}$ under the element of the T-duality
group which shifts the B-field.
For the large volume torus (or $K3$) we have the following relation between
the numbers of branes $(Q_4,Q_2,Q_0)$ and the parameters of the 
$1+5$-dimensional Yang-Mills:
\begin{equation}
(Q_4, Q_2, -Q_0)=\trace \exp \left({1\over 2\pi i} F\right)
\end{equation}
(For the large volume torus, the Yang-Mills theory lives on $D5$ branes,
and the rank of the gauge group is $Q_4$. The sign of $Q_0$ may be understood
in the following way. Let us consider the $D1D5$ system wrapped on $K3$. 
On one hand, to get BPS configuration we should have $Q_0Q_4>0$.
On the other hand, D-strings should become anti-self-dual gauge fields on 
$K3$, so that $\int\trace \left({1\over 2\pi i}F\right)^2<0$).

The six-dimensional string coupling constant is related to the
ten-dimensional $g_{str}$:
\begin{equation}\label{kappa}
\kappa={g_{str}\over \sqrt{V}}
\end{equation}
This $\kappa$ is invariant under T duality. 

\section{An instanton in the presence of a string: an example.}
Here we want to prove that the configuration consisting 
of the black string with the charge $Z$ and the D-instanton
with the charge vector $\chi$ preserves $1\over 8$ of
supersymmetry, if $(\chi\cdot Z)=0$, $(\chi\cdot P_{W_4}Z)=0$
and $||\chi||^2\geq 0$.   
We will consider the configuration consisting of the black 
string with $Q_2=0$ (only $D1$ and $D5$, no $D3$), and the
instanton obtained by wrapping the $D1$-string on the holomorphic 
cycle $\alpha$ in ${\bf T}^4$. It is enough to consider this 
configuration, because any other pair of orthogonal vectors 
$\chi$ and $Z$ can be brought to this one by $SO(4,4,{\bf R})$ 
transformations (we do not need this transformations to be defined
over ${\bf Z}$, because we want just to understand whether supersymmetry
is preserved, and this is the property of the classical supergravity
solution).

Let us temporarily turn off the $B$ field.
The Euclidean $Dp$-brane stretched along the coordinates 
$y^1,\ldots, y^{p+1}$ preserves the combination of supercharges,
satisfying the condition:
\begin{equation}
\Gamma_1\cdots\Gamma_{p+1} \epsilon_L=i\epsilon_R
\end{equation}
For example, for the $D_{-1}$ instanton, $\epsilon_L=i\epsilon_R$
\cite{GreenGutperle}. Let us denote $x^0$ and $x^5$ coordinates 
parallel to the black string, and $x^1,\ldots,x^4$ coordinates in the
torus. The supersymmetries preserved by the $D1/D5$ system are:
\begin{equation}\label{SusyOfDoneDfive}
\begin{array}{c}
\Gamma_0\Gamma_5 \epsilon_L=i\epsilon_R \\
\Gamma_0\Gamma_5\Gamma_1\cdots\Gamma_4\epsilon_L=i\epsilon_R
\end{array}
\end{equation}
Suppose that the  cycle $\alpha$ on which we wrap our D-string to get an 
instanton is holomorphic in some complex structure of the torus.
The corresponding K\"ahler form $\omega_{\bf R}$ can act on spinors
(being contracted with gamma-matrices). We will denote the corresponding
operator $\homegr$. The supersymmetry preserved by the instanton is 
selected by the condition that $\epsilon_L$ is chiral in the tangent 
space to ${\bf T}^4$ and $\epsilon_R$ is:
\begin{equation}
\homegr \epsilon_L=i\epsilon_R
\end{equation}
This equation, together with (\ref{SusyOfDoneDfive}) implies that 
$\epsilon_L$ is chiral in the tangent space of ${\bf T}^4$ and also
subject to the condition $\homegr\epsilon_L=\Gamma_0\Gamma_5\epsilon_L$.
Then, $\epsilon_R$ is expressed in terms of $\epsilon_L$ by one of
the equations (\ref{SusyOfDoneDfive}). 
We see that our background preserves $1\over 8$ of the supersymmetry.

Our background was quite generic except that we have put $B=0$.
To turn on the $B$ field, we use the subgroup 
$SO(3,3,{\bf R})\subset SO(4,4,{\bf R})$ which preserves both
$Z=(Q_4,0,Q_0)$ and $\chi=(0,\alpha,0)$.  This groups contains
rotations in the plane generated by $(Q_4,0,-Q_0)$ and $(0,\beta,0)$, where
$\beta\in H^2({\bf T}^4,{\bf R})$ is orthogonal to $\alpha$. One can see
that infinitesimal rotations in such a plane transform our background 
with zero $B$
field to the background with the infinitesimal $B$ field $\delta B$ 
satisfying the condition $(\delta B\cdot \alpha)=0$ and otherwise
arbitrary. This means, that there is one and only one condition for
the background with nonzero $B$ to be connected to the background with zero
$B$ by the rotation orthogonal to both $Z$ and $\chi$. This condition
is $(Z\cdot P_{W_4}\chi)=0$.

\end{document}